\documentclass[preprint]{aastex}
\pdfoutput=1
\usepackage{epsfig}
\usepackage{morefloats}
\usepackage{natbib}
\usepackage{amsmath}
\usepackage{multirow}
\usepackage[inline,draft,nomargin,author=]{fixme}
\fxusetheme{colorsig}
\usepackage{algpseudocode}
\usepackage{algorithmicx}
\usepackage{algorithm}

\makeatletter
\let\@dates\relax
\makeatother

\newcommand{\vc}[1]{ {\bf{#1}} }
\newcommand{\grad}[1]{ \vc{\nabla } #1 }
\newcommand{\dv}[1]{ \vc{\nabla } \cdot #1\ }

\begin {document}

\title{A Hybrid Advection Scheme for Conserving Angular Momentum on a Refined Cartesian Mesh}

\author{Zachary D. Byerly\altaffilmark{1}, Bryce Adelstein-Lelbach\altaffilmark{2}, Joel E. Tohline\altaffilmark{1,2} and Dominic C. Marcello\altaffilmark{1,2}}

\altaffiltext{1}{Department of Physics and Astronomy,
 Louisiana State University, Baton Rouge, LA 70803}
\altaffiltext{2}{Center for Computation \& Technology,
 Louisiana State University, Baton Rouge, LA 70803}

\begin{abstract}
We test a new ``hybrid'' scheme for simulating dynamical fluid flows in which cylindrical components of the momentum are advected across a rotating Cartesian coordinate mesh. This hybrid scheme allows us to conserve angular momentum to machine precision while capitalizing on the advantages offered by a Cartesian mesh, such as a straightforward implementation of mesh refinement. Our test focuses on measuring the real and imaginary parts of the eigenfrequency of unstable axisymmetric modes that naturally arise in massless polytropic tori having a range of different aspect ratios, and quantifying the uncertainty in these measurements. Our measured eigenfrequencies show good agreement with the results obtained from the linear stability analysis of \citet{kojima1986} and from nonlinear hydrodynamic simulations performed on a cylindrical coordinate mesh by \citet{WTH1994}.  When compared against results conducted with a traditional Cartesian advection scheme, the hybrid scheme achieves qualitative convergence at the same or, in some cases, much lower grid resolutions and conserves angular momentum to a much higher degree of precision.  As a result, this hybrid scheme is much better suited for simulating astrophysical fluid flows, such as accretion disks and mass-transferring binary systems.
\end {abstract}

\section{Introduction}

\subsection{Context}

Binary star systems, especially those containing compact components, are of great current interest in astrophysics. Binaries containing white dwarf components, in particular, are quite common because white dwarfs represent the most frequent endpoint for stellar evolution. Even double white dwarf (DWD) binaries, which are formed through common envelope evolution, are estimated to number $\sim 2.5 \times 10^8$ in our Galaxy \citep{Nelemans2001}. When gravitational radiation drives these binaries to a semi-detached state, these mass transferring systems are thought to be progenitors to both type Ia supernovae \citep{Geier2007,Rosswog2009,Fryer2010,Schaefer2012} and to hydrogen-poor R Coronae Borealis (RCB) stars \citep{Webbink1984,Staff2012}. \citet{Yungelson2004} review many possible evolutionary paths of binary systems. \citet{Tylenda2011} point to the evolution of V1309 Sco as an example of a system in which an actual merger has been witnessed observationally.

Our desire is to employ computational fluid techniques to model, in a self-consistent manner and to a high degree of accuracy, mass-transfer in a wide variety of interacting binary star systems over hundreds, if not thousands, of orbits. Examples of our efforts, to date, include \citet{DSouza2006}, \citet{Motl2007}, \citet{Even2009}, and \citet{Marcello2012}. Related work by other groups includes \citet{Benz1990}, \citet{Fryer2006,Fryer2010,Fryer2012}, \citet{Yoon2007}, and \citet{Raskin2012}. Such a capability would allow us not only to better understand the behavior of binaries that are dynamically unstable toward merger or tidal disruption of the donor, but also to examine how spin-orbit coupling -- for example, the exchange of angular momentum between the donor star and a disk surrounding the accretor -- facilitates dynamical stability and leads to long phases of quasi-steady mass transfer. With such a tool we could simulate how slow accretion can bring an initially sub-Chandrasekhar-mass accretor to the brink of critical collapse; how a transition from sub- to super-Eddington accretion rates affects common-envelope development and evolution; and the steady-state structure of mass-transferring AM CVn type binaries.

In simulating these systems numerically, a faithful representation of the flow will be achieved only if the grid resolution is sufficiently high across a range of dynamically interesting flow regions.  These regions can vary in structure as well as in identity over time, so the grid needs to adapt accordingly.  For example, even when only considering double degenerate binaries, the smallest length scales (surface layers of both stars, scale height of the disk, and fluid flow through the L1 Lagrange point) can be tiny compared to the binary separation. At the other extreme, an envelope consisting of an optically thick atmosphere engulfing both stars can develop shortly after accretion begins, quickly filling the original computational domain.  This ``common envelope'' structure may expand to a size many times larger than the binary separation.  Adaptive mesh refinement (AMR) techniques can be called upon to provide an appropriately high degree of spatial resolution in various, as well as in time-varying, regions of the flow. Astrophysical simulation codes that employ AMR include FLASH \citep{Fryxell2000}, ZEUS \citep{Hayes2006}, PLUTO \citep{Mignone2007}, and Scorpio, recently developed by Marcello et al. (2014, in preparation). AMR techniques can be straightforwardly implemented on a Cartesian mesh but are more difficult to employ across curvilinear grids. Despite this difficulty, AMR has been implemented on both spherical and cylindrical coordinate grids \citep{Fryxell2000}.

On the other hand, advection schemes implemented on Cartesian meshes are most naturally designed to conserve linear momentum, rather than angular momentum.  Because binary evolutions can be faithfully followed through hundreds of orbits only if the simulation conserves angular momentum to a high degree of accuracy, in the past we have chosen to use a cylindrical computational grid, which more naturally facilitates conservation of orbital angular momentum. However, even a cylindrical grid does not match the symmetry of each individual binary component. And in the absence of a mesh refinement capability we have not had the full freedom to distribute resolution where it is needed. 

Alternative curvilinear meshes designed to address some of the above outlined shortcomings have been used in previous disk and torus simulations. For example, \citet{Zink2008} discuss a multi-patch technique meant for spherically symmetric or axisymmetric simulations, and \citet{Fragile2009} introduce a ``patched-sphere'' mesh similar to the multi-patch technique.  But, as with a cylindrical coordinate grid, such schemes do not easily accommodate mesh refinement techniques.  Ultimately, \citet{Zink2008} say that Cartesian mesh-refinement is likely better suited for problems like binary mergers.  Historically, therefore, it has been difficult to achieve both high -- and adaptive -- spatial resolution while at the same time achieving a high degree of angular momentum conservation. The hybrid advection scheme designed by \citet{Call2010} and implemented here, allows us to have our cake and eat it, too. It facilitates conservation of angular momentum to machine accuracy on a refined Cartesian mesh.  

 \citet{Mignone2012} state that their method, implemented in the PLUTO code, allows for conserving angular momentum on a Cartesian mesh to machine precision.  However, it appears as though this only applies to local ``shearing-box'' models. Their scheme breaks down the azimuthal fluid velocity into two pieces, an average plus a residual term.  The average fluid velocity is handled in a linear step by moving the fluid in the direction of the orbital motion.  The residual velocity is then handled in the standard way.  Only the residual portion is subject to the Courant condition, so this leads to larger time steps, and the apparent motion of the fluid through the grid is much smaller, leading to less numerical dissipation.  However, this method requires that the average angular motion be parallel to one of the coordinate bases describing the grid and that the relevant grid direction use periodic boundary conditions.  This is not the case for a global simulation performed on a Cartesian mesh, as we are doing in this work.  The method implemented in PLUTO would be a very suitable choice for simulating axisymmetric tori as we have done in this work.  It would not, however, provide a significant advantage for binary mass-transfer simulations, even if performed on a cylindrical grid, as the fluid is largely rotating with a uniform angular velocity.

While our new method will allow us to conserve angular momentum at a level that is necessary to faithfully follow interacting binary simulations through thousands of orbits, it does not ease the computational burden of simulating the large number of time steps needed for such a simulation. In particular, the Courant limit on the size of individual time steps remains an impediment. But by facilitating the straightforward implementation of AMR, our new scheme simplifies the task of load balancing and thereby enhances a code's ability to efficiently and more fully use the capabilities of massively parallel computers, allowing simulations to be carried through thousands of orbits. As our discussion in \S \ref{octopus} indicates, the port of our new hybrid scheme to Octopus is yet another step toward achieving this goal.

\subsection{Overview of this Work}

In this paper we demonstrate the utility of the hybrid scheme by focusing on a quantitative analysis of nonaxisymmetric, dynamical instabilities that arise spontaneously in Papaloizou-Pringle tori \citep{PP1984}, hereafter referred to as PP tori.  Each PP torus is a non-self-gravitating, differentially rotating, geometrically thick, axisymmetric disk in orbit about a central point mass.  Its  internal structure is defined by a balance between gas pressure gradients and gradients in the effective potential.  The vertical thickness of the disk/torus relative to its radial extent is determined by the choice of the polytropic index for the gas and an initial angular momentum distribution.  (See \S2 for details.)  These configurations are suitable for demonstrating the capabilities of our hybrid scheme because:  They each have a simple analytically definable initial state; while each initial state is axisymmetric, the system is unstable to the development of nonaxisymmetric structure, hence, its evolution has a fully three-dimensional character; the eigenvector of the most unstable mode for each chosen initial configuration, while not known analytically, should be well defined and its measured properties -- for example, its complex eigenfrequency -- should be reproducible and independent of the specific numerical scheme that is used to perform the dynamical simulation. At the same time, the PP torus provides a good test for hydro codes such as ours because of the challenges it provides.  A Cartesian mesh is not ideally suited for the initially axisymmetric torus problem, and thus gives the hybrid scheme an opportunity to prove its worth, for example, by partially overcoming the spurious $m=4$ modes that are excited by the structure of an underlying Cartesian grid. 

Each of our simulations is carried out on a rotating and refined Cartesian grid.  The hydrodynamic code we are using (see \S\S 2.2-2.3) has AMR capabilities, but for simplicity we have chosen not to activate the AMR feature.  Instead, for each simulation the volume of the grid that is occupied by the initial torus is resolved using a time-invariant, fixed level of refinement (LOR).  The effect of grid resolution is assessed by repeating individual simulations several times, using a different (fixed) number of refinement levels.  (Typically, we employ 4, 5, or 6 LOR.)  In addition, for each initial state and for each specified LOR, the dynamical evolution is carried out using two separate advection schemes:  (1) A traditional ``Cartesian'' scheme in which the $x$, $y$, and $z$ components of the linear momentum are advected across the refined Cartesian grid; and (2) our new ``hybrid'' scheme in which radial momentum and angular momentum -- instead of the $x$ and $y$ components of the momentum -- are advected across the refined Cartesian grid.  In total, results from 23 simulations are reported here.

For each simulation, the real and imaginary components of the complex eigenfrequency of the fastest growing unstable mode are measured and, as appropriate, compared with previously published results from the nonlinear hydrodynamic simulations performed on a uniformly zoned cylindrical mesh by \citet*{WTH1994} and from the linear stability analysis presented by \citet{kojima1986}.  In an effort to eliminate any subjective bias that might be introduced into the measurement of these eigenfrequencies and, at the same time, to facilitate future efforts to reproduce our results, we introduce a mathematically prescriptive method for quantifying both the value of and uncertainty in each eigenfrequency measurement.  In doing this we are able to meaningfully assess the performance of our new hybrid advection scheme, relative to the performance of the traditional Cartesian advection scheme.  We show that qualitative convergence is achieved with the hybrid scheme at the same, or sometimes at significantly lower, grid resolutions.

At the same time, we show that the hybrid scheme allows conservation of the system's total angular momentum to machine accuracy.  As explained above, this is a highly desirable feature that is not possible to achieve using a familiar Cartesian advection scheme. This is perhaps the most significant attribute of our hybrid scheme.  Historically, the expectation has been that angular momentum conservation can be achieved when modeling an astrophysical disk or binary system only if one adopts a coordinate grid -- for example, cylindrical or spherical coordinates -- whose underlying basis vectors accommodate the curvilinear features of the flow.  Our hybrid scheme facilitates conservation of angular momentum on a Cartesian grid.

The hydrocode that has been used to carry out the primary set of simulations reported in this paper employs OpenMP to enable multiple execution threads within a single, multi-core compute node.  All of the models in our primary set of simulations -- totaling 22 in number and using up to 6 LOR -- fit within a single node of our high-performance computing system.  In \S3 of this paper we present results from one simulation that was conducted on a rotating Cartesian grid with 7 LOR. This single simulation was run using Octopus, a closely aligned hydrocode built on
top of High Performance ParalleX (HPX), a newly emerging parallel runtime system. 

\section{Methods}
\subsection{Initial Models}
As has already been stated, the initial axisymmetric equilibrium models used in this study were all geometrically thick (uniform specific angular momentum) massless tori with structures as derived by \citet{PP1984}. There have been a number of published studies of the nonaxisymmetric stability of these massless tori.  Some of these are linear stability analyses \citep{PP1984,kojima1986,Frank1988}, while others are nonlinear hydrodynamics simulations, either in two dimensions \citep{Hawley1987, Hawley1990}, or in full three dimensions \citep{Zurek1986, Hawley1990, WTH1994}. Our initial models have been chosen to provide a direct comparison with the linear stability analysis of \citet{kojima1986} and the nonlinear hydrodynamic results of \citet*[hereafter WTH]{WTH1994}.  

We generated the same sequence of seven initial models used by WTH, as detailed here in Table \ref{table:modeltable}; Figure \ref{kojima_profiles} displays the cross-sectional surface of each.  The models in this sequence vary in the ratio of the inner radius $R_-$ to the outer radius $R_+$.

Following PP, we assume a polytropic equation of state for the fluid,
\begin{equation}
\label{EOS}
p(\rho) = K\rho^{1+ \frac{1}{n}},
\end{equation}
where, $n$ is the polytropic index and $K$ is the polytropic constant, and we impose a power-law rotation profile given by,
\begin{equation}
\label{RotationProfile}
\Omega(R) = \Omega_0\left(\frac{R_0}{R}\right)^q,
\end{equation}
where $\Omega_0 = \Omega(R_0)$ is the angular velocity in the equatorial plane at the radius of pressure maximum, $R_0$.  Throughout this work we assume $n=3$ and $q=2$ (uniform specific angular momentum). As PP have shown, a solution of the hydrostatic balance equation,
\begin{equation}
\label{Equilibrium}
-\frac{1}{\rho}\nabla p - \nabla\Phi_\mathrm{eff} = 0,
\end{equation}
where,
\begin{equation}
\Phi_\mathrm{eff} \equiv - \frac{GM_\mathrm{pt}}{(R^2 + Z^2)^{1/2}} + \frac{1}{2}\Omega_0^2(R)R^2 , 
\end{equation}
yields the following axisymmetric equilibrium density distribution:
\begin{figure}[ht]
\centering
\includegraphics[width=\linewidth]{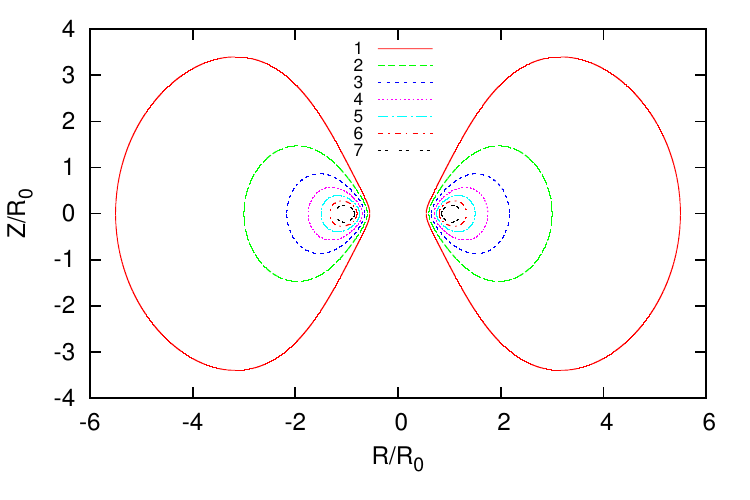}
\caption{Meridional-plane cross sections of the seven different axisymmetric equilibrium disks evolved in this paper, listed in Table \ref{table:modeltable}. Each disk has a $n=3$ polytropic index and a uniform specific angular momentum ($q=2$).  Lengths have been normalized to the radius of pressure maximum, $R_0$. }
\label{kojima_profiles}
\end{figure}
\begin{equation}
\label{DensityProfile}
\rho_\mathrm{A}(R,Z) = 
\left(\frac{GM_\mathrm{pt}}{4K}\right)^3
\left[
\frac{1}{(R^2 + Z^2)^{1/2}} - \frac{R_-}{R^2(1+R_-/R_+)} - \frac{1}{(R_+-R_-)}
\right]
 ^3,
\end{equation}
where, $M_\mathrm{pt}$ is the specified mass of the central point mass. In our simulations, we normalize all lengths to the outer edge of the torus, such that $R_+ = 1.0$, and vary $R_-$. (For illustration purposes, Figure \ref{kojima_profiles} is handled differently.)  The mass $M_\mathrm{pt}$ is chosen such that $\rho_\mathrm{max} = \rho_\mathrm{A}(R_0,0) = 1.0$.

\begin{deluxetable}{clll}
\tablecolumns{4}
\tablewidth{0pt}
\tablecaption{Torus Geometries \label{table:modeltable}}
\tablehead{\colhead{Model} & \colhead{$\frac{R_-}{R_+}$} & \colhead{$\frac{R_0}{R_+}$} & \colhead{$\frac{R_-}{R_0}$}}
\startdata
1 & 0.1 & 0.1818 & 0.55 \\
2 & 0.2 & 0.3333 & 0.60 \\
3 & 0.3 & 0.4615 & 0.65 \\
4 & 0.4 & 0.5714 & 0.70 \\
5 & 0.5 & 0.6666 & 0.75 \\
6 & 0.6 & 0.7500 & 0.80 \\
7 & 0.7 & 0.8235 & 0.85 \\
\enddata
\end{deluxetable}

Before each model was introduced into the hydrocode, the initial axisymmetric density distribution was perturbed by a single-mode ``kick'',
$$ \rho(R,Z,\theta) = \rho_A(R,Z)[1 + a\cos(m_\mathrm{kick}\theta)], $$
where, $m_\mathrm{kick}$ is the specific azimuthal mode that we choose to excite, and $a = 10^{-2}$. No velocity perturbation was introduced.

\subsection{Hydrodynamics Code}

The perturbed models are evolved on a rotating Cartesian grid using Scorpio (Marcello et al. 2014, in preparation) with modifications as detailed here.  We enforce a polytropic equation of state that is evolved with the same polytropic index, $n=3$, that was used to construct the initial models, and a mass density floor of $10^{-10}\rho_\mathrm{max}$. Since the PP torus is massless, we evolve the system in a time-independent gravitational potential given by a central object of mass $M_\mathrm{pt}$.  For more code details, please see Appendix \ref{appendix_code}.

\begin{figure}[ht]
\centering
\includegraphics[angle=270,scale=0.17,keepaspectratio=true]{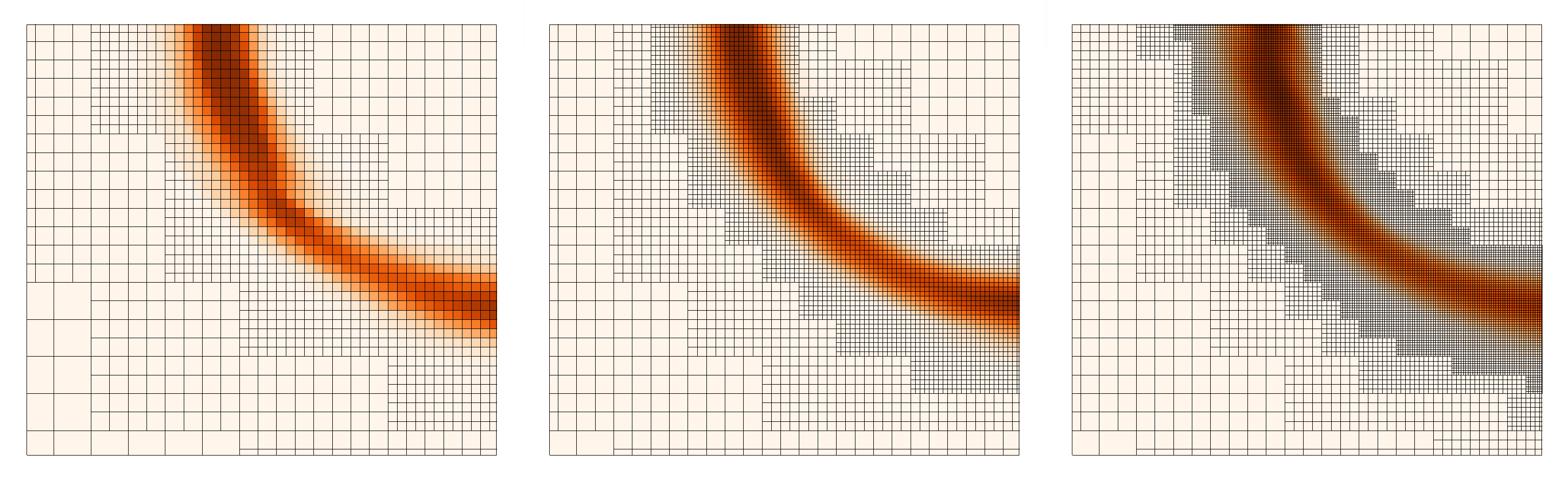}
\caption{A comparison of the mesh structure of a $R_-/R_+=0.7$ torus simulation taken from three different resolutions.  From top to bottom are 4, 5, and 6 LOR. All simulations run using Cartesian momentum advection.  4 LOR simulation shown at $t= 1.5 t_{orb}$, 5 LOR simulation shown at $t= 2.0 t_{orb}$, and 6 LOR simulation shown at $t= 2.8 t_{orb}$. }
\label{rescompare}
\end{figure}

This work presents simulations run at different resolutions.  Because we use an octree-based grid, it is convenient to speak in terms of the number of levels of refinement of the coarsest $8\times 8\times 8$ grid.  Each additional level of refinement doubles the resolution in the most refined region. A simulation run at ``$N$'' LOR has a maximally refined grid spacing $\Delta x = (x_\mathrm{max} - x_\mathrm{min})/(8\cdot 2^N)$, where $x_\mathrm{max}$ and $x_\mathrm{min}$ are the maximum and minimum extents of the simulation domain, respectively.  For the purposes of this study, we use a fixed mesh refinement based on the geometry of the torus, rather than an adaptive mesh. As illustrated in Figure \ref{rescompare}, the grid is fully refined throughout the volume occupied by the initially axisymmetric torus. 

Table \ref{table:resolution} gives three different numbers for each model at each LOR in an effort to fully reveal the grid properties. For each model, the top row repeats the ratio of the inner radius to the outer radius, $R_-/R_+$, shown in Table \ref{table:modeltable}; a measure of the slenderness of the torus.  The next three rows give three different numbers for 4 LOR: the number (in millions) of fully-refined (smallest grid spacing) zones, the number of fully-refined (leaf) subgrids (each containing $8\times 8 \times 8$ zones), and the ratio of the radius at pressure maximum to the finest grid spacing, $R_0/\Delta x$.  The next nine rows give the same set of numbers for 5, 6 and 7 LOR.  The first two numbers are related by a factor of $8^3 = 512$, so Table \ref{table:resolution} gives two potential metrics for quantifying how well a torus is resolved.  The first is simply the total number of fully-refined zones inside of the torus, which decreases with increasing $R_-/R_+$; while the ratio  $R_0/\Delta x$ increases with $R_-/R_+$. These two metrics seem to contradict each other. The first (number of fully-refined zones) suggests that at any given LOR, a fatter (lower $R_-/R_+$) torus is better resolved, while the second ($R_0/\Delta x$) would indicate that a skinnier (higher $R_-/R_+$) torus is better resolved. This illustrates that comparing resolution between different models is not straightforward.  

It seems clear that determining whether or not the resolution used to simulate a particular model is sufficient must be done on a model-by-model basis. \cite{Hawley2013} distinguish between true numerical convergence and ``qualitative'' convergence.  In our hydrodynamic simulations, we do not expect to be able to achieve true numerical convergence, characterized by errors in the solution approaching zero.  Rather, we strive to achieve qualitative convergence, a condition that is obtained when physically important macroscopic quantities do not change by a significant amount with an increase in resolution.  However, one must be cautious when using this criterion, as the lack of change from one resolution to another may only be an indication that neither resolution is resolving some important physical process.  In the case of our rather idealized toroidal models, the physically relevant dynamic processes that determine the eigenvectors of unstable modes should be resolved sufficiently well that this will not be a problem.

\begin{deluxetable}{cccccccccc}
\tablecolumns{10}
\tabletypesize{\small}
\tablewidth{0pt}
\tablecaption{Resolution\tablenotemark{a} \label{table:resolution}}
\tablehead{    &           & \multicolumn{7}{c}{$R_-/R_+$} \\
LOR &  & 0.1 & 0.2 & 0.3 & 0.4 & 0.5 & 0.6 & 0.7}
\startdata
 & zones ($\times10^6$) & \nodata & \nodata & 0.303 & \nodata & \nodata & \nodata & 0.106 \\
4 & subgrids & \nodata & \nodata & 592 & \nodata & \nodata & \nodata & 208 \\
 & $R_0/\Delta x$ & \nodata & \nodata & 19.23 & \nodata & \nodata & \nodata & 34.31 \\
\hline
 & zones ($\times10^6$) & 2.43    & 2.07    & 1.64  & 1.38  & 0.987  & 0.844 & 0.668 \\
5 & subgrids            & 4,744   & 4,040   & 3,208 & 2,696 & 1,928 & 1,648 & 1,304 \\
 & $R_0/\Delta x$       & 15.15   & 27.78   & 38.46 & 47.62 & 55.56 & 62.50 & 68.63 \\
\hline
 & zones ($\times10^6$) & \nodata & \nodata & 10.73 & \nodata & \nodata & \nodata & 3.232 \\
6 & subgrids & \nodata & \nodata & 20,952 & \nodata & \nodata & \nodata & 6,312 \\
 & $R_0/\Delta x$ & \nodata & \nodata & 78.22 & \nodata & \nodata & \nodata & 139.58 \\
\hline
 & zones ($\times10^6$) & \nodata & \nodata & \nodata & \nodata & \nodata & \nodata & 19.31 \\
7 & subgrids & \nodata & \nodata & \nodata & \nodata & \nodata & \nodata & 37,712 \\
 & $R_0/\Delta x$ & \nodata & \nodata & \nodata & \nodata & \nodata & \nodata & 279.16 \\
\enddata
\tablenotetext{a}{Information is shown only for simulations performed here.}
\end{deluxetable}

\subsection{Description of Hybrid Angular Momentum Conservation Scheme}

We evolve the following two coupled fluid equations,
\begin{equation}
\frac{\partial}{\partial t}\rho + \dv{\rho \vc{u}} = 0,
\end{equation}
and
\begin{equation}
\label{momentum_eq}
\frac{\partial}{\partial t} (\rho \vc{v})  + \dv{ (\rho \vc{v}  \vc{u})  } = - \grad{ p } -\rho\grad{ \Phi },
\end{equation}
where $\rho$ is the mass density, $p$ is the gas pressure, both $\vc{v}$ and $\vc{u}$ identify the same fluid velocity field (that is, $\vc{v} = \vc{u}$), and $\Phi$ is the gravitational potential generated by a central point mass, $M_\mathrm{pt}$, specifically,
\begin{equation}
\Phi = -\frac{GM_\mathrm{pt}}{(x^2+y^2+z^2)^{1/2}}.
\end{equation}
We use two different variables to represent the same velocity field to emphasize that, following \cite{Call2010}, we have the freedom to choose different coordinate bases for each of the velocity terms that appear in the dyadic tensor product, $\vc{v}\vc{u}$, in the nonlinear advection term of equation (\ref{momentum_eq}). See Appendix \ref{appendix_eqs} for further elaboration.

We evolve the fluid with the same polytropic equation of state, equation (\ref{EOS}), given in \S2.1.  Because we are using this polytropic equation of state, there is no need to evolve an energy equation.

When rewriting the ``momentum conservation'' equation (\ref{momentum_eq}) in terms of three orthogonal vector components, we begin by identifying two familiar sets of equations: When advecting Cartesian momentum components -- $s_x \equiv \rho v_x$, $s_y \equiv \rho v_y$, $s_z \equiv \rho v_z$ -- we start with,
\begin{eqnarray}
\frac{\partial}{\partial t}s_x + \nabla\cdot{(s_x\vc{u})} = - \hat{\vc{e}}_x\cdot\grad{p} - \hat{\vc{e}}_x\cdot\grad{\Phi},\\
\frac{\partial}{\partial t}s_y + \nabla\cdot{(s_y\vc{u})} = - \hat{\vc{e}}_y\cdot\grad{p} - \hat{\vc{e}}_y\cdot\grad{\Phi},\\
\frac{\partial}{\partial t}s_z + \nabla\cdot{(s_z\vc{u})} = - \hat{\vc{e}}_z\cdot\grad{p} - \hat{\vc{e}}_z\cdot\grad{\Phi};
\end{eqnarray}
and, when advecting cylindrical momentum components -- $s_R \equiv \rho v_R, \ell_z \equiv R\rho v_\varphi, s_z \equiv \rho v_z$ -- the $z$-component is identical to the Cartesian case but the other two orthogonal components are,
\begin{eqnarray}
\label{nn}
\frac{\partial}{\partial t}s_R + \nabla\cdot{(s_R\vc{u})} &=& -\hat{\vc{e}}_R\cdot\grad{p} -\hat{\vc{e}}_R\cdot\rho\grad{\Phi} + \frac{\ell_z^2}{\rho R^3},\\
\label{mm}
\frac{\partial}{\partial t}\ell_z + \nabla\cdot{(\ell_z\vc{u})} &=& -R\hat{\vc{e}}_\varphi\cdot\grad{p} -R\hat{\vc{e}}_\varphi\cdot\rho\grad{\Phi},
\end{eqnarray}
where, $R \equiv (x^2+y^2)^{1/2}$.

These familiar sets of equations are morphed into the sets of equations used in our hybrid scheme by recognizing several things. First, as is demonstrated in Appendix \ref{appendix_eqs}, in the divergence term of all five identified momentum component equations, we can immediately replace $\vc{u}$ by the velocity field as viewed from a frame of reference that is rotating at angular frequency, $\Omega_0$, namely,
\begin{equation}
\vc{u}' = \vc{u} - \hat{\vc{e}}_\varphi R\Omega_0,
\end{equation}
because $\dv{(\hat{\vc{e}}_\varphi R\Omega_0)} = 0$, that is, because the velocity field introduced by the frame rotation is divergence free. All of the other elements of the five component equations remain unchanged when $\vc{u}$ is replaced by $\vc{u}'$ -- in particular, all five advected quantities, $s_x, s_y, s_z, s_R,$ and $\ell_z,$ still refer to components of the inertial-frame momentum (or angular-momentum) density.  Second, when advecting Cartesian components of the momentum across a rotating grid, an additional source term,
\begin{equation}
\vc{S}_\textrm{rot} = -\vc{\Omega_0}\times(\rho\vc{u}) = \Omega_0s_y\vc{\hat{e}}_x-\Omega_0s_x\vc{\hat{e}}_y,
\end{equation}
must be inserted on the right-hand side to account for the time-varying orientation of the Cartesian unit vectors. These modifications permit us to rewrite all three Cartesian components of the momentum conservation equation in the form that we have used for this project:

\begin{eqnarray}
\frac{\partial}{\partial t}s_x + \nabla\cdot{(s_x\vc{u}')} &=& - \frac{\partial}{\partial x}p - \rho\frac{\partial}{\partial x}\Phi +\Omega_0s_y,\\
\frac{\partial}{\partial t}s_y + \nabla\cdot{(s_y\vc{u}')} &=& - \frac{\partial}{\partial y}p - \rho\frac{\partial}{\partial y}\Phi -\Omega_0s_x,\\
\frac{\partial}{\partial t}s_z + \nabla\cdot{(s_z\vc{u}')} &=& - \frac{\partial}{\partial z}p - \rho\frac{\partial}{\partial z}\Phi.
\end{eqnarray}

Third, we note that an evaluation of the advection term that appears on the left-hand-side of each component of the momentum equation, which is generically of the form,
\begin{equation}
\dv{(\Psi \vc{u}')},
\end{equation}
requires an assessment of the divergence of the three-dimensional flow field at each location on the computational grid. But, in practice, it shouldn't matter whether this ``assessment'' is done on a Cartesian mesh or on a cylindrical mesh (or on any of a multitude of other mesh choices); the result should be the determination of the proper scalar value at every point on the chosen computational grid. So, although the familiar form of the set of equations governing the time-rate-of-change of the cylindrical components of the momentum, presented above, was derived with the implicit assumption that each term would be evaluated on a cylindrical coordinate mesh, we can just as well evaluate the advection term on a Cartesian mesh. This only requires that the divergence operator and the ``transport'' velocity, $\vc{u}'$, be handled in exactly the same manner as they are handled when evaluating advection in the Cartesian set of equations. In the hybrid scheme being presented here, all simulations are conducted on a Cartesian mesh so, in all cases, the divergence operator and the transport velocity are broken down into Cartesian components before the advection term is evaluated. 

Finally, because a Cartesian mesh is being adopted, the gradient operator on the right-hand-side of each component of the momentum equation is also explicitly broken down into its Cartesian components. This means that, for our hybrid scheme, the right-hand-sides of equations (\ref{nn}) and (\ref{mm}) incorporate the operator projections,
\begin{eqnarray}
\hat{\vc{e}}_R\cdot\nabla &=&  \left[\hat{i}\left(\frac{x}{R}\right)+\hat{j}\left(\frac{y}{R}\right)\right]\cdot\nabla    = \frac{x}{R}\frac{\partial}{\partial x} + \frac{y}{R}\frac{\partial}{\partial y},\\
\hat{\vc{e}}_\varphi\cdot\nabla &=& \left[\hat{j}\left(\frac{x}{R}\right)-\hat{i}\left(\frac{y}{R}\right)\right]\cdot\nabla = \frac{y}{R}\frac{\partial}{\partial x} - \frac{x}{R}\frac{\partial}{\partial y}.
\end{eqnarray}

With all of these recognitions in hand, in our hybrid scheme the three components of the cylindrical momentum equations are,
\begin{eqnarray}
\label{cyl_cart_rot_R}
\frac{\partial}{\partial t}{s_R} + \dv{( s_R \vc{u'} )} &=&  -\frac{1}{R}\left(x\frac{\partial}{\partial x} + y\frac{\partial}{\partial y}\right)p - \frac{\rho}{R}\left(x\frac{\partial}{\partial x} + y\frac{\partial}{\partial y}\right)\Phi + \frac{\ell_z^2}{\rho  R^3},\\
\label{cyl_cart_rot_lz}
\frac{\partial}{\partial t}{\ell_z} + \dv{( \ell_z \vc{u'}) } &=& \left(y\frac{\partial}{\partial x} - x\frac{\partial}{\partial y}\right)p + \rho\left(y\frac{\partial}{\partial x} - x\frac{\partial}{\partial y}\right)\Phi,\\
\frac{\partial}{\partial t}{s_x} + \dv{( s_x \vc{u'}) } &=& - \frac{\partial}{\partial z} p - \rho\frac{\partial}{\partial z}\Phi.
\end{eqnarray}
As discussed by \cite*{Call2010}, it is noteworthy that the right-hand-side of the hybrid-scheme equation that governs transport (and conservation) of angular momentum does not contain a Coriolis term. This is because $\ell_z$, the quantity being advected and tracked, is the angular momentum density as measured in the \textit{inertial} frame of reference. As is demonstrated in \S A.4 of Appendix A, a Coriolis term arises if the equation is written in a form where the quantity being advected is the \textit{rotating-frame} angular momentum density. This equation, which contains a Coriolis term, is more familiar to the astrophysics community -- see, for example, \cite*{Norman1980} and \cite*{New1997}. But, for purposes of angular momentum conservation, we consider it to be far preferable to adopt a version of the equation in which the velocity does not explicitly appear in the source term.

Finally, we note that $s_R$ and $\ell_z$ can be straightforwardly expressed in terms of Cartesian components of $\vc{u}$ or $\vc{u}'$. Specifically, remembering that $\vc{u}=\vc{v}$, 
\begin{eqnarray}
s_R &\equiv& \rho v_R = \frac{\rho}{R}(xu_x+yu_y) = \frac{\rho}{R}(xu'_x+yu'_y),\\
\ell_z &\equiv& R\rho v_\varphi = \rho(xu_y-yu_x) = \rho(xu'_y-yu'_x)+\rho\Omega_0(x^2+y^2).
\end{eqnarray}

\subsection{Postprocessing}
\begin{figure}
\centering
\includegraphics[height=\textheight,width=\textwidth,keepaspectratio=true]{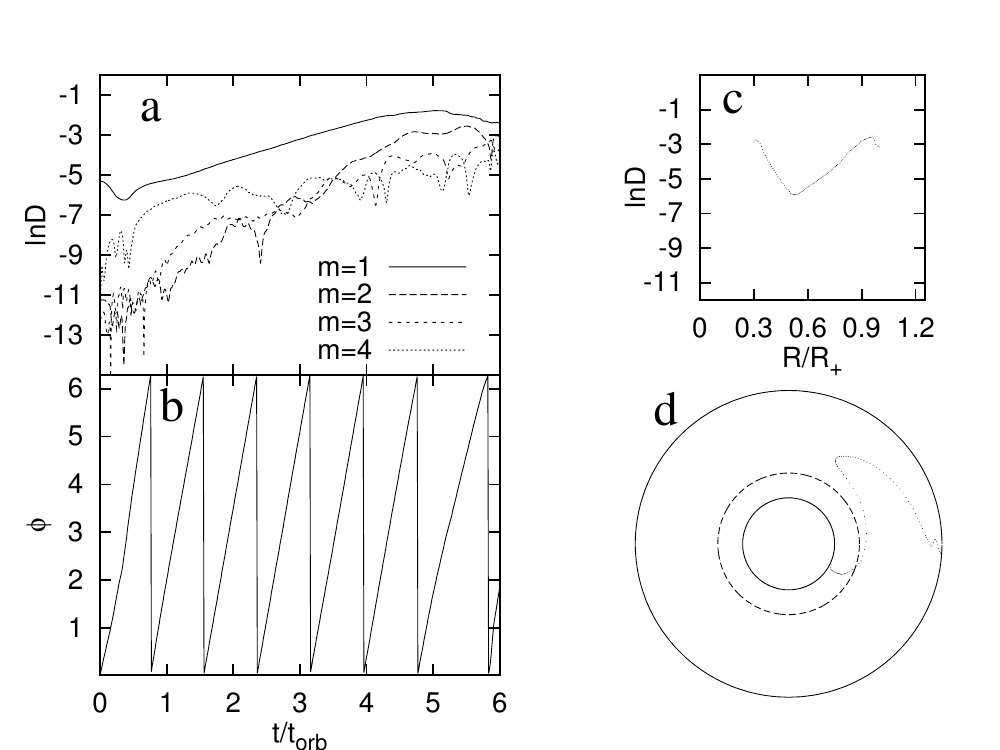}
\caption{Data from the Cartesian advection simulation of Model 3 ($R_-/R_+ = 0.3$) is used to demonstrate four diagnostic diagrams, providing a direct comparison to Figure 2 in \cite{WTH1994}.  (a) A ``$D_m - t$'' diagram showing the Fourier amplitude of modes $m=1,2,3,4$ at the radius of pressure maximum in the equatorial plane. (b) A ``$\phi_m-t$'' diagram showing the phase angle of the $m=1$ mode, again at the radius of pressure maximum in the equatorial plane.  (c) A ``$D_m - r$'' diagram showing the amplitude of the $m=1$ mode as a function of radius in the equatorial plane at time $t=2.5\ t_\mathrm{orb}$. (d) A ``$\phi_m-r$'' diagram showing the azimuthal location of the density maximum ($m=1$) as a function of radius in the equatorial plane at time $t=2.5\ t_\mathrm{orb}$.}
\label{4_panel}
\end{figure}

We want to quantify small deviations from the initial axisymmetric density distribution, which can be described by,
\begin{equation}
\frac{\delta\rho}{\rho} = f(R,Z) e^{-i[\omega t-m\theta]},
\end{equation}
where, $m$ is the azimuthal mode number and $\omega$ is a complex frequency.
In order to do this, we describe the mass density in the system in terms of a discrete Fourier series,
\begin{equation}
  \rho(J,L,K,t) = \frac{1}{2}c_0(J,K,t) + \sum\limits_{m=1}^{L_\mathrm{max}} c_m(J,K,t)\cos\left[m\frac{2\pi L}{L_\mathrm{max}}+\phi_m(J,K,t)\right],
\end{equation}
where $J$ is the radial index, $K$ is the vertical index, and $L$ is the azimuthal index.
For each Z we use a two-dimensional third-degree bivariate spline technique (\cite{bivariate}, as implemented in scipy.interpolate.RectBivariateSpline) to interpolate mass density values from the Cartesian grid ($x,y$) to a polar coordinate grid ($R,\theta$). We then employ a Fourier transform to determine the coefficient of the discrete Fourier series as follows:
\begin{eqnarray*}
a_m(J,K) &=& \frac{2}{\pi} \sum^{L_\mathrm{max}}_{L=1} \rho(J,L,K)\cos(mL\delta\theta)\delta\theta,\\
b_m(J,K) &=& \frac{2}{\pi} \sum^{L_\mathrm{max}}_{L=1} \rho(J,L,K)\sin(mL\delta\theta)\delta\theta,\\
c_m(J,K) &=& [a_m(J,K)^2 + b_m(J,K)^2]^{1/2},\\
\phi_m(J,K) &=& \arctan[-b_m(J,K)/a_m(J,K)],\\
D_m &\equiv& \frac{c_m}{\sum^{L_\mathrm{max}}_{i=0}c_i}.
\end{eqnarray*}
After running each simulation and plotting the time variation of $D_m$ and $\phi_m$ for various modes, the components of the complex frequency $\omega$ for a dynamically growing eigenmode of the system can be written as,
\begin{eqnarray*}
\mathrm{Re}(\omega) = \frac{d\phi_m}{dt},\\
\mathrm{Im}(\omega) = \frac{d\ln c_m}{dt}.
\end{eqnarray*}
In order to compare with \cite{kojima1986}, we use the quantities,
\begin{equation}
y_1 \equiv [\mathrm{Re}(\omega)/\Omega_0-m] = \left[\frac{1}{\Omega_0}\frac{d\phi_m}{dt}-m\right],
\end{equation}
and
\begin{equation}
y_2 \equiv \mathrm{Im}(\omega)/\Omega_0 = \left[\frac{1}{\Omega_0}\frac{d\ln c_m}{dt}\right],
\end{equation}
to describe the properties of the unstable modes.
When plotting $\ln D_m$ vs time (a ``$D_m - t$'' diagram, as shown, for example, in Figure \ref{4_panel}a), where $t$ is normalized to $t_\mathrm{orb} \equiv 2\pi/\Omega_0$, the value of $y_2$ is the slope of the line divided by $2\pi$. Similarly, $y_1$ is straightforwardly obtained when plotting $\phi_m$
versus time (a ``$\phi_m - t$'' diagram, Figure \ref{4_panel}b) by taking the period measured from the
graph, dividing by $2\pi$, and then subtracting $m$.  Note that $\phi_m$ must be
plotted in the inertial frame. Here we usually will present results only at the radius of pressure maximum,
$R_0$, and in the equatorial plane.  In order to gain a better understanding of the structure of these fluctuations
at different radii, however, we can perform the same Fourier transform at many different
radii along the equatorial plane.  We can then obtain a more complete picture of the structure of the unstable eigenvector by plotting, for one particular point
in time, the amplitude of a single Fourier mode as a function of radius (a
``$D_m - r$'' diagram, Figure \ref{4_panel}c), and the phase angle of a single Fourier mode as a
function of radius (a ``$\phi_m - r$'' diagram, Figure \ref{4_panel}d).

\subsection{Quantifying Results}
We have devised a formulaic method for quantifying both the growth rate and the quality of the measurement.  The idea is that an analysis of simulations that display long, relatively quiet periods of exponential growth should produce much smaller error bars than the analysis of simulations that display only short, noisy segments of exponential growth. 
A formulaic method of this type minimizes human judgment and helps ensure that the measurements are reproducible. 

\begin{figure}
\centering
\includegraphics[scale=1.8]{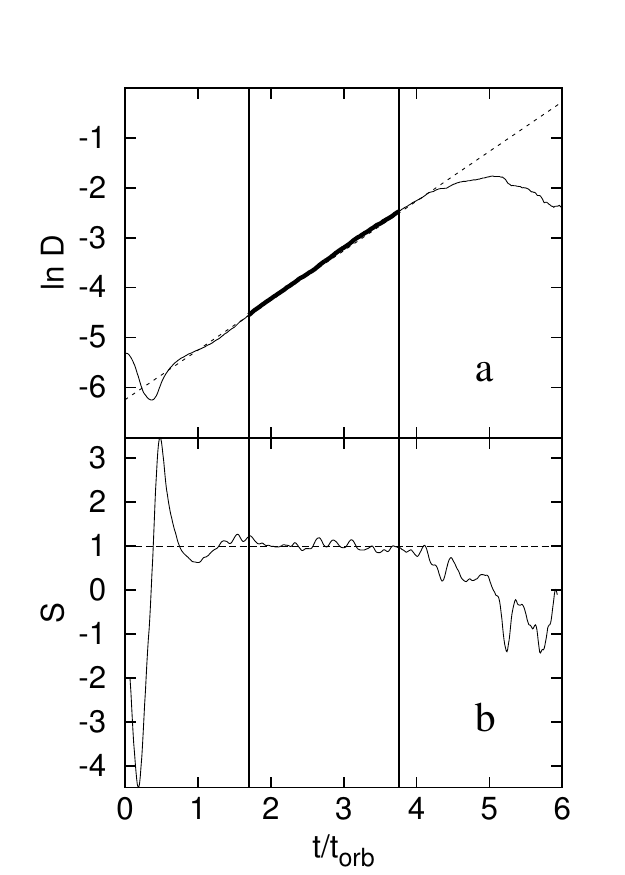}
\caption{Data from a simulation of Model 3 ($R_-/R_+ = 0.3$) using the hybrid scheme and 5 LOR. (a) Shows a $D_m - t$ plot, with vertical lines marking the starting and ending points of the region used to determine $\bar{S}$; the portion of the curve used to measure this slope is shown in bold.  A dashed line with a slope equal to the $\bar{S}$ is also shown.  (b) Shows the windowed slope measurement, $S(t)$, with vertical lines marking the starting and ending points of the region used to measure $\bar{S}$.  The horizontal dashed line identifies the measured slope. }
\label{postprocessing1}
\end{figure}

\begin{figure}
\centering
\includegraphics[scale=2]{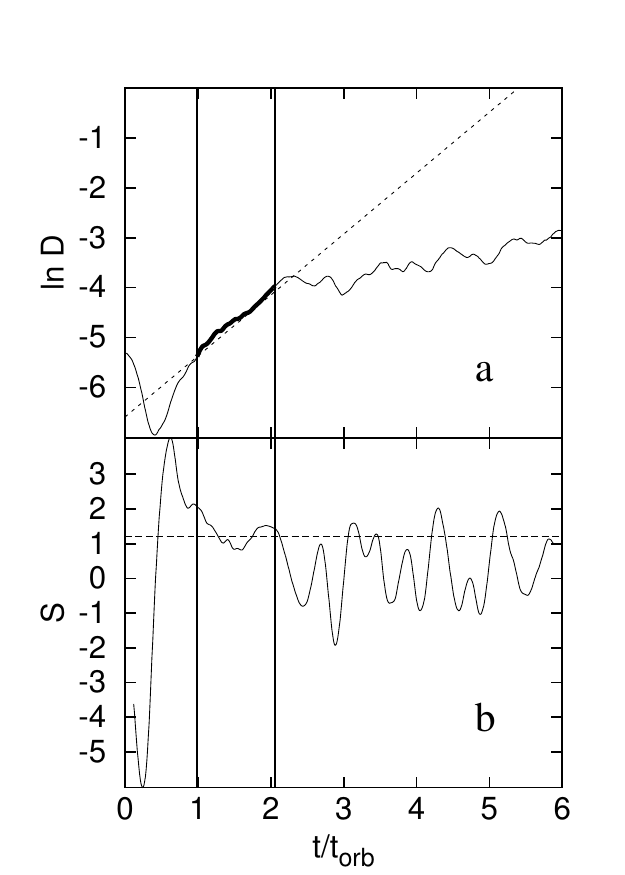}
\caption{Same as Figure \ref{postprocessing1}, but from a simulation of Model 7 ($R_-/R_+ = 0.7$) using the Cartesian momentum advection scheme and 5 LOR.}
\label{postprocessing2}
\end{figure}

As is illustrated in Figures \ref{postprocessing1}a and \ref{postprocessing2}a we begin the analysis by generating a $D_m - t$ plot from the results of each simulation. (A full suite of $D_m - t$ plots from our 5 LOR evolutions is shown in Figure \ref{postprocessing}.) This plotted $D_m - t$ curve contains 100 individual data points for each orbit. Starting at $t = t_\mathrm{orb}/8$, with a window of width $t_\mathrm{orb}/4$ centered on that time, a linear regression is used to determine the slope, $S$.  This window is moved continuously across the $D_m - t$ plot to generate a set of data, $S(t)$, as shown in Figures \ref{postprocessing1}b and \ref{postprocessing2}b.  

The relatively flat portion of the Figure \ref{postprocessing1}b $S(t)$ curve -- bounded by the two vertical solid lines at $t_\mathrm{start} = 1.7\ t_\mathrm{orb}$ and $t_\mathrm{end} = 3.76\ t_\mathrm{orb}$ -- corresponds to the period of relatively clean exponential growth in the Figure \ref{postprocessing1}a $D_m - t$ plot.  In order to determine the best portion of the simulation over which to measure a slope of the  $D_m - t$ graph, that is, in order to determine the best time boundaries, $t_\mathrm{start}$ and $t_\mathrm{end}$, we first define the error bar size as,
\begin{equation}
\delta_{y_2} = \frac{\sqrt{\frac{1}{i_\mathrm{end}-i_\mathrm{start}}\displaystyle\sum\limits_{i=i_\mathrm{start}}^{i_\mathrm{end}}(S_i - \bar{S})^2}}{t_\mathrm{end}-t_\mathrm{start}},
\end{equation}
where $i_\mathrm{start}$ and $i_\mathrm{end}$ are the starting and ending points of the selected section respectively, and $S_i$ is the value of $S(t)$ at each time $t_i$. Every allowable combination of $i_\mathrm{start}$ and $i_\mathrm{end}$ is considered (as long as $t_\mathrm{end} - t_\mathrm{start} \geq 1.0\ t_\mathrm{orb}$), and the best time boundaries are defined by the interval which generates the smallest error bar, $\delta_{y_2}|_\mathrm{min}$. We then define the measured slope to be the mean of $S(t)$ in the selected region,
\begin{equation}
2\pi y_2 = \bar{S} = \frac{1}{i_\mathrm{end}-i_\mathrm{start}}\displaystyle\sum\limits_{i=i_\mathrm{start}}^{i_\mathrm{end}} S_i.
\end{equation}
The magnitude of $\delta_{y_2}|_\mathrm{min}$ is not meaningful by itself, but it can be used to compare the relative quality between two separately measured slopes. 

Figures \ref{postprocessing1} and \ref{postprocessing2} show examples of $D_m - t$ plots (top) with their corresponding $S(t)$ plots (bottom).  The dashed lines shown in both the $D_m - t$ and $S(t)$ plots represent the measured slope, $\bar{S}$, for the relevant model simulation. Figure \ref{postprocessing1} shows data from a simulation that generated a relatively small $\delta_{y_2}|_\mathrm{min}$. In contrast, Figure \ref{postprocessing2} shows data that generated a larger $\delta_{y_2}|_\mathrm{min}$.  Notice that, when compared to Figure \ref{postprocessing1}, the region of exponential growth is shorter and $S(t)$ is much less constant in the selected region identified in Figure \ref{postprocessing2}.

Additionally, we sought to measure the $y_1$ parameter in a way that minimized human interaction.  The portion of the $\phi_m - t$ graph that corresponds with the region selected above ($t_\mathrm{start}$ to $t_\mathrm{end}$) was fitted to a line using a least-squares method.  The slope of this line is used to compute $y_1$, and the residuals of this fit are then used to compute $\delta_{y_1}$.

\section{Results}

\begin{figure}
\centering
\includegraphics[angle=90,scale=1.3]{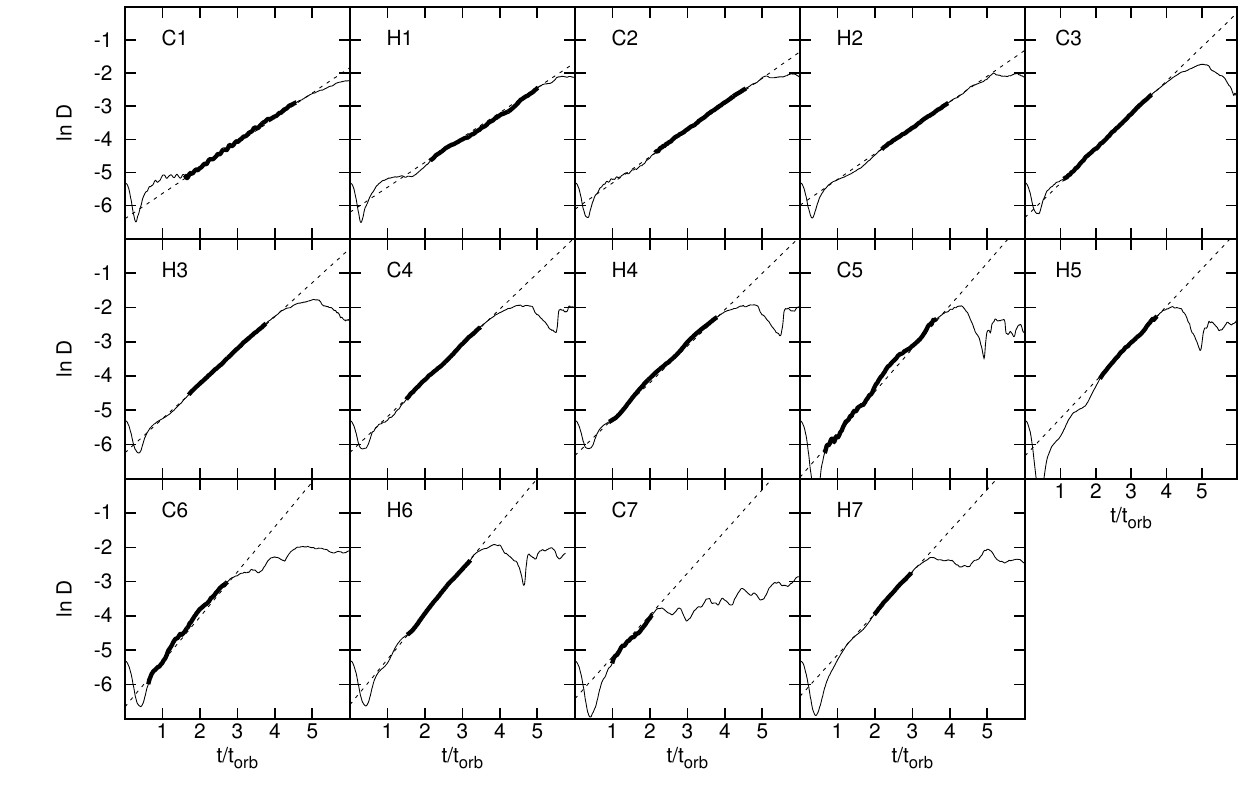}
\caption{A series of $D_m - t$ plots from every initial model listed in Table \ref{table:modeltable} labeled with either ``H'' or ``C'' for simulations using the hybrid scheme or the Cartesian momentum advection scheme, respectively, followed by the model number. All simulations were performed using 5 LOR. In each case the portion of the plot used to measure $\bar{S}$ is highlighted in bold, and a dashed line with the measured slope is also plotted.}
\label{postprocessing}
\end{figure}

Figure \ref{postprocessing} shows $D_m - t$ plots resulting from 14 separate simulations, all run at 5 LOR, with the selected linear portion of the plot in bold, and a dashed line representing the linear fit.  Table \ref{table:results} contains a separate row of data corresponding to each of these 14 simulations. Column 1 identifies the initial model configuration as specified in Table \ref{table:modeltable}, column 2 identifies the azimuthal mode perturbation that we introduced at the start of each run, and column 3 indicates whether the simulation was carried out using a standard Cartesian (C) advection scheme or our new hybrid (H) scheme. The measured slope, $\bar{S}$, associated value of $y_2$, and error bar, $\delta_{y_2}|_\mathrm{min}$, corresponding to the Figure \ref{postprocessing} plots are tabulated in columns 4, 5, and 6, respectively, of Table \ref{table:results}.  From the $\phi_m - t$ plot of each simulation (not shown), we measured $d\phi/dt$, $y_1$, and $\delta y_1$ and recorded the results in columns 7, 8, and 9, respectively, of Table \ref{table:results}.

\begin{figure}
\centering
\includegraphics[scale=2]{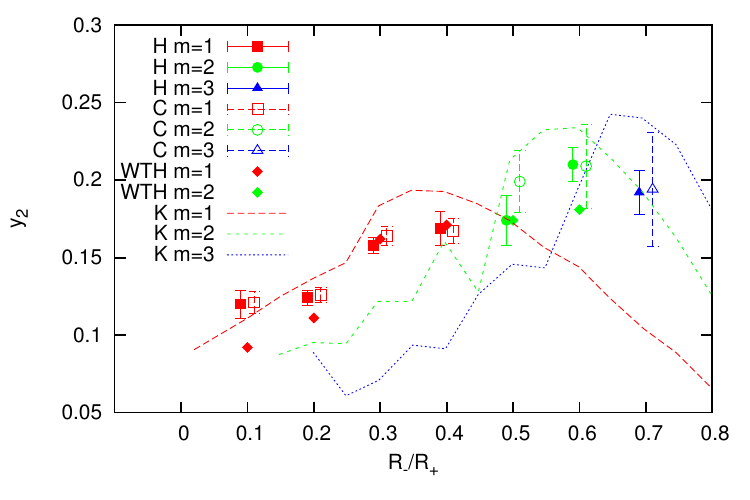}
\caption{Comparison of the imaginary component ($y_2$) of the eigenfrequency of various unstable modes between this work and \citet{WTH1994} and \citet{kojima1986}. The red, green, and blue dashed curves connect discreet points from Kojima's linear analysis for $m=1,2,$ and 3 respectively. The points marked with red boxes, green circles, and blue triangles show values measured in this work for $m=1,2,$ and 3 respectively; open symbols represent the Cartesian momentum advection scheme and filled symbols represent the hybrid scheme. Red and green diamonds represent $m=1$ and 2 results published in WTH, respectively. Our measured growth rates show good agreement with both previous studies.  As described in the text, error bars on data points from this work represent the relative quality of measurements.}
\label{y_2comparison}
\end{figure}

\begin{figure}
\centering
\includegraphics[scale=2]{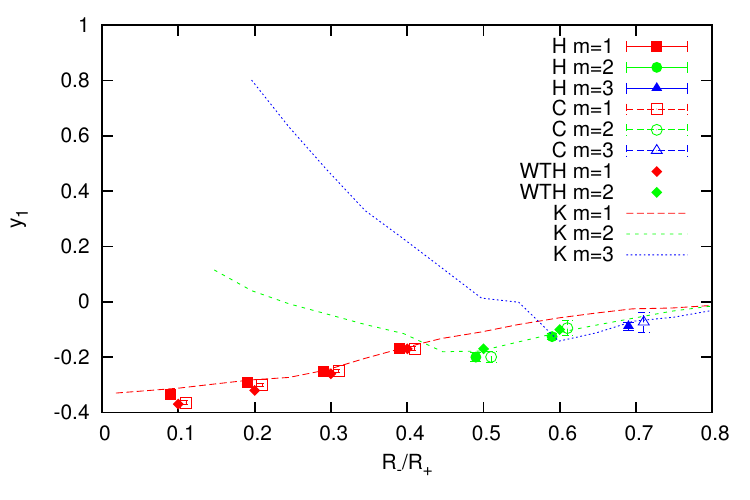}
\caption{As in Figure \ref{y_2comparison}, but showing the real component ($y_1$) of the eigenfrequency of various unstable modes. Our measured frequencies show good agreement with both previous studies.}
\label{y_1comparison}
\end{figure}

When we apply the post-processing algorithm to the results of our simulations, we see a good match with both the results of Kojima and WTH, although error bars were not provided in either of these earlier works.  In Figures \ref{y_2comparison} and \ref{y_1comparison} we plot, respectively, the imaginary ($y_2$) and real ($y_1$) components of the eigenfrequency as determined from our simulations, from WTH, and from the linear analysis of Kojima.  The three dashed curves connect discreet points from the Kojima linear analysis.  Because the modes are completely uncoupled in his simulations, Kojima is able to measure growth rates for all modes over most of the range of $R_+/R_-$ that we are interested in.  Results from WTH are marked by solid diamonds. Our current results, shown as symbols with error bars, represent the $m=1$ growth rate for the four fattest tori (the points marked by red boxes), the $m=2$ growth rate for the next 2 models (points marked by green circles), and the $m=3$ growth rate for the slimmest torus (points marked by blue triangles).  Results from simulations performed with the traditional Cartesian advection scheme are indicated by open symbols while results using our hybrid scheme are indicated by filled symbols. For each model these symbols are separated horizontally on the graph purely for visibility; both runs were started from identical initial states.

While comparison to Kojima's results provides a good sanity check for our results, they do not really provide an ideal solution to our problem.  Kojima's linear analysis problem has different boundary conditions than the nonlinear simulations performed here and by WTH.  Kojima assumes that the surface of the torus remains fixed in space, whereas the hydrocodes allow the torus surface to distort.  In the linear analysis, the individual modes do not couple, whereas in the nonlinear hydrocode they do.  Thus we do not expect that our results will converge to Kojima's, no matter how much resolution we use. 

In addition to the information displayed in Figure \ref{y_2comparison} highlighting the fastest growing modes, we observe in all simulations an unphysical development of $m=4$ distortions.  This is undoubtedly due to the 4-fold symmetry of the underlying Cartesian grid.  As is illustrated more fully below, in all simulations the time-evolutionary development of $m=4$ distortions is strongly resolution dependent. The simulations using the hybrid scheme show less unphysical $m=4$ growth than do the simulations advecting Cartesian momentum.

\begin{figure}
\centering
\includegraphics[scale=1.8]{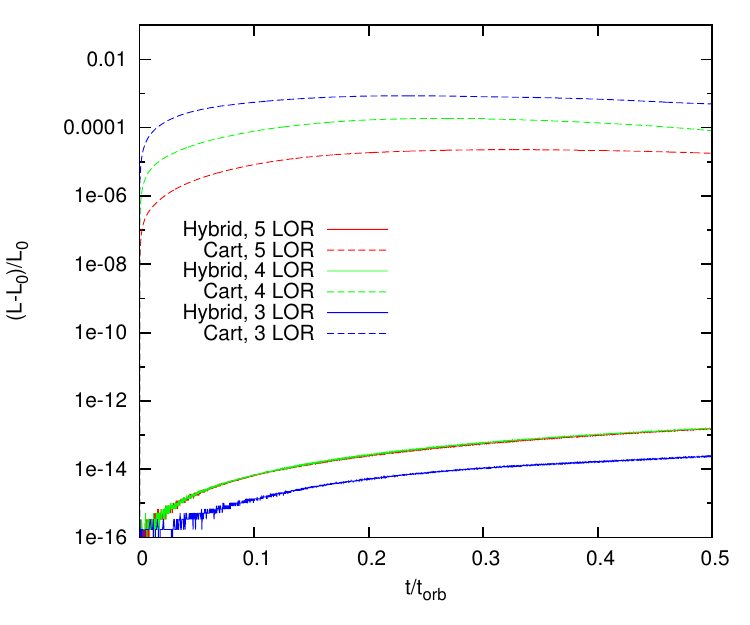}
\caption{
The accumulated change in Model 7's total angular momentum $(L-L_0)$, measured relative to its initial value, $L_0$, is shown as a function of time, $t/t_\mathrm{orb}$, for 6 different simulations -- 3 different levels of refinement and using Cartesian momentum advection (dashed curves) or the hybrid scheme (solid curves). In the case of both schemes, the red, green, and blue curves show data from simulations conducted with 5, 4, and 3 LOR respectively.  When using the Cartesian momentum advection scheme, the level of angular momentum conservation shows clear resolution dependence. The hybrid scheme conserves angular momentum at a level set by machine truncation error.}
\label{angmom_comparison}
\end{figure}

Figure \ref{angmom_comparison} shows the global conservation of angular momentum for a $R_-/R_+=0.7$ torus at 3 different LOR.  These curves show the difference between the initial total angular momentum, $L_0$, and the angular momentum, $L$, at time $t$, divided by the initial total angular momentum.  The red, green, and blue dashed curves show data from the Cartesian momentum advection scheme at 5, 4, and 3 LOR respectively; the red, green, and blue solid curves show data from the hybrid scheme at 5, 4, and 3 LOR respectively. The simulations run with the hybrid scheme (solid curves) show non-conservation at levels $\Delta L/ L_0 \lesssim 10^{-13}$ due only to machine truncation error.  The hybrid scheme simulation at 3 LOR (solid blue curve) seems to display better conservation than the 4 and 5 LOR simulations (overlapping solid green and red curves), however this is only because the 3 LOR simulation takes far fewer time steps than the higher resolution simulations and thus accumulates less error due to truncation.  The simulations run with the Cartesian momentum advection scheme display resolution dependent global angular momentum conservation, at levels of $\Delta L/ L_0 \approx 10^{-5}$, $10^{-4}$, and $10^{-3}$, for 5, 4, and 3 LOR, respectively.  Examining the resolution dependence of the conservation of angular momentum by the Cartesian momentum advection scheme can give us a measure of the convergence of the hydro scheme.  The difference between 3 and 4 LOR means that $dx$ is cut in half, and this corresponds to an order-of-magnitude difference in the level of conservation, as shown in Figure \ref{angmom_comparison}.  The same is true again for the difference between 4 and 5 LOR. This corresponds to $~O(dx^{2.5})$ level of convergence, consistent with the expectations of the scheme used by Scorpio (see Appendix B).

The hybrid scheme is able to conserve angular momentum to a very high precision and significantly outperform the Cartesian advection scheme, on this test problem, largely because the imposed gravitational field is purely radial. No source terms due to gravity appear in the azimuthal component of the momentum equation but source terms due to gravity do appear in all three components of the Cartesian momentum equation.  One might ask which scheme performs better if the primary concern is conservation of linear momentum. Our chosen problem is not well suited for testing conservation of linear momentum precisely because source terms due to gravity appear in all three components of the Cartesian momentum equation. We concede that the outcome would very likely have been reversed -- that is, the standard Cartesian advection scheme would have significantly outperformed the hybrid scheme, as described here -- had we chosen a problem in which the gradient in the gravitational potential was zero in one or more of the directions defined by the Cartesian mesh. But the beauty of the more general hybrid scheme, as described by \cite*{Call2010}, is that the grid geometry can be picked independently of the specific problem while the components of the momentum vector that are advected can be easily modified, in response to the structure of the underlying force field. This even facilitates switching back and forth between, say, Cartesian and cylindrical momentum components during a simulation.

\begin{figure}
\centering
\includegraphics[scale=1.6]{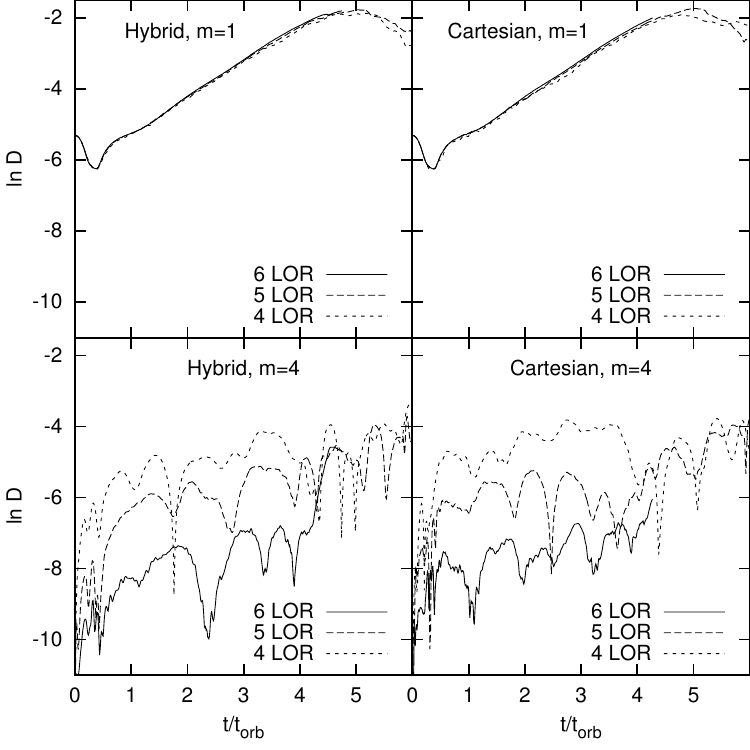}
\caption{Data from Model 3 ($R_-/R_+ = 0.3$) simulations. Each of the six panels shows the $D_m - t$ plot for $m=1$ (solid curve) and $m=4$ (dashed curve) obtained from, as labeled, either the hybrid scheme or the Cartesian advection scheme for 4, 5, or 6 LOR.  While the time-dependent growth of the unstable $m=1$ mode is very similar in all cases, the amplitude of $m=4$ appears to be strongly resolution dependent, reflecting the 4-fold symmetry of the Cartesian grid structure.}
\label{model_3_compare_LOR}
\end{figure}

\begin{figure}
\centering
\includegraphics[scale=2]{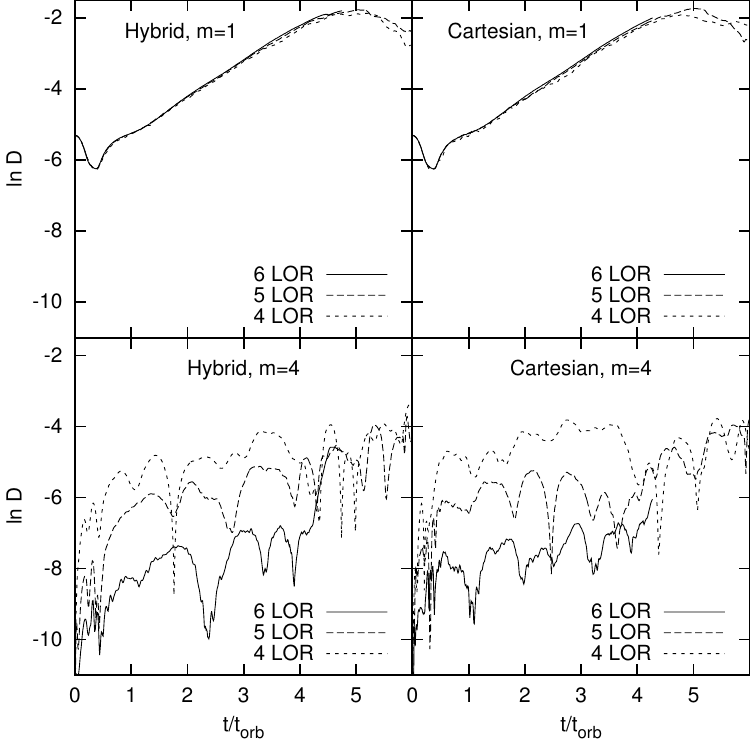}
\caption{The same information as shown in Figure \ref{model_3_compare_LOR}, but grouped differently. Each frame shows all three levels of refinement for either $m=1$ (top row) or $m=4$ (bottom row), using either the hybrid scheme (left column) or Cartesian momentum advection scheme (right column). The two upper panels ($m=1$) show that both the hybrid and Cartesian seem to converge, as the lower resolution simulations display noisier $D_m - t$ plots.  The lower two panels show how the amplitude of $m=4$ fluctuations decreases dramatically with increasing grid resolution.}
\label{model_3_compare_mode}
\end{figure}

\begin{figure}
\centering
\includegraphics[scale=1.4]{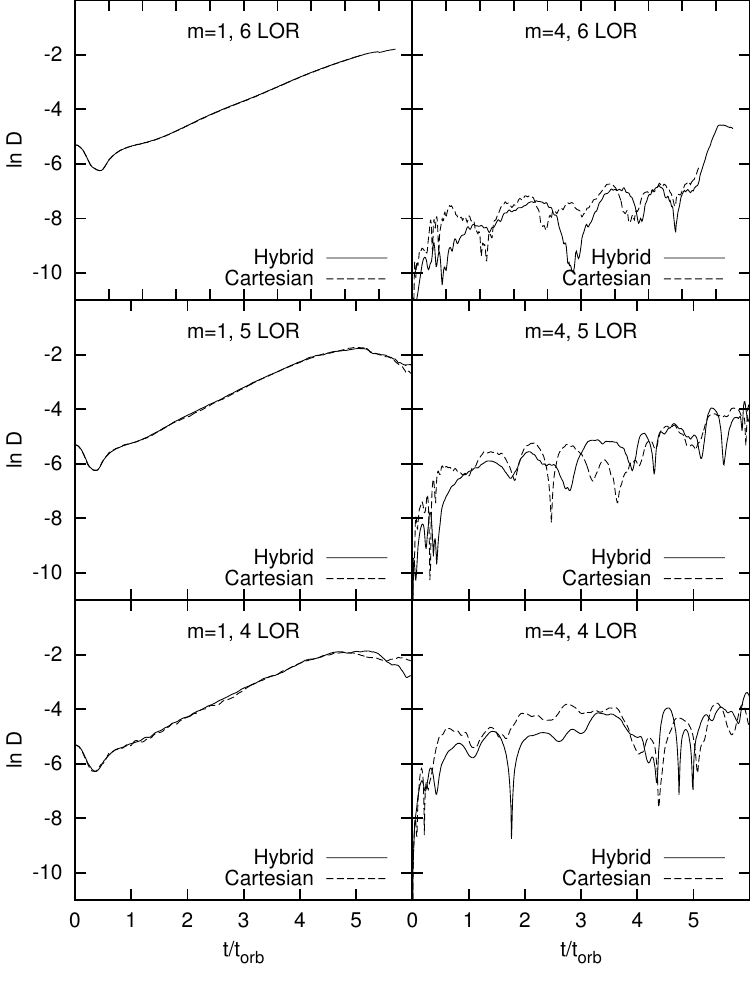}
\caption{The same information as shown in Figure \ref{model_3_compare_LOR}, but grouped differently.  Each frame compares hybrid (solid curve) to Cartesian (dashed curve) momentum advection schemes, for either $m=1$ (left column) or $m=4$ (right column), and at 4, 5, and 6 levels of refinement (bottom, middle, and top rows, respectively). The left column ($m=1$) illustrates how the difference between the Cartesian and hybrid scheme plots differs less with increasing resolution, suggesting that they are both converging to the same answer. The right column ($m=4$) shows that, at each specified resolution, the hybrid scheme generally produces slightly lower amplitude $m=4$ distortions than the Cartesian advection scheme.}
\label{model_3_compare3}
\end{figure}

\begin{figure}
\centering
\includegraphics[scale=1.8]{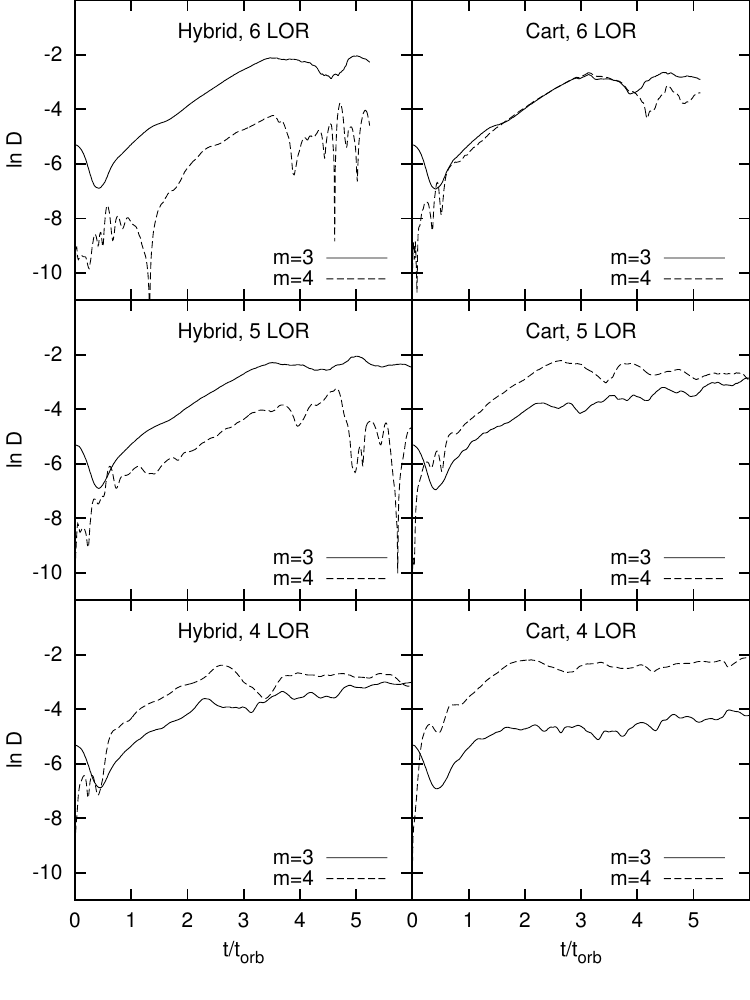}
\caption{Data from Model 7 ($R_-/R_+ = 0.7$) simulations. Each of the six panels shows the $D_m - t$ plot for $m=3$ (solid curve) and $m=4$ (dashed curve) obtained from, as labeled, either the hybrid scheme or the Cartesian advection scheme for 4, 5, or 6 LOR.}
\label{model_7_compare_LOR}
\end{figure}

\begin{figure}
\centering
\includegraphics[scale=2]{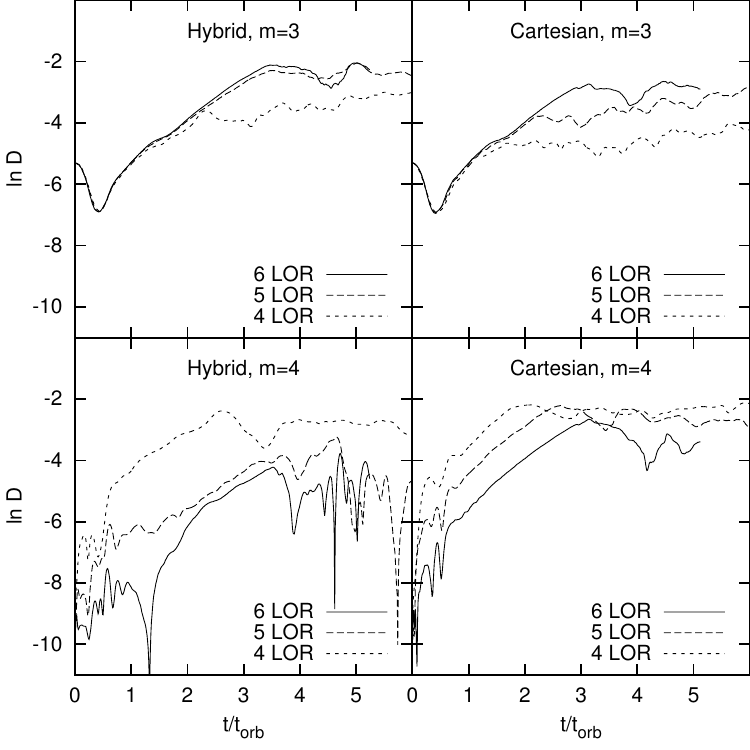}
\caption{The same information as shown in Figure \ref{model_7_compare_LOR}, but grouped differently. Each frame shows all three levels of refinement for either $m=3$ (top row) or $m=4$ (bottom row), using either the hybrid scheme (left column) or Cartesian momentum advection scheme (right column).}
\label{model_7_compare_mode}
\end{figure}

\begin{figure}
\centering
\includegraphics[scale=1.6]{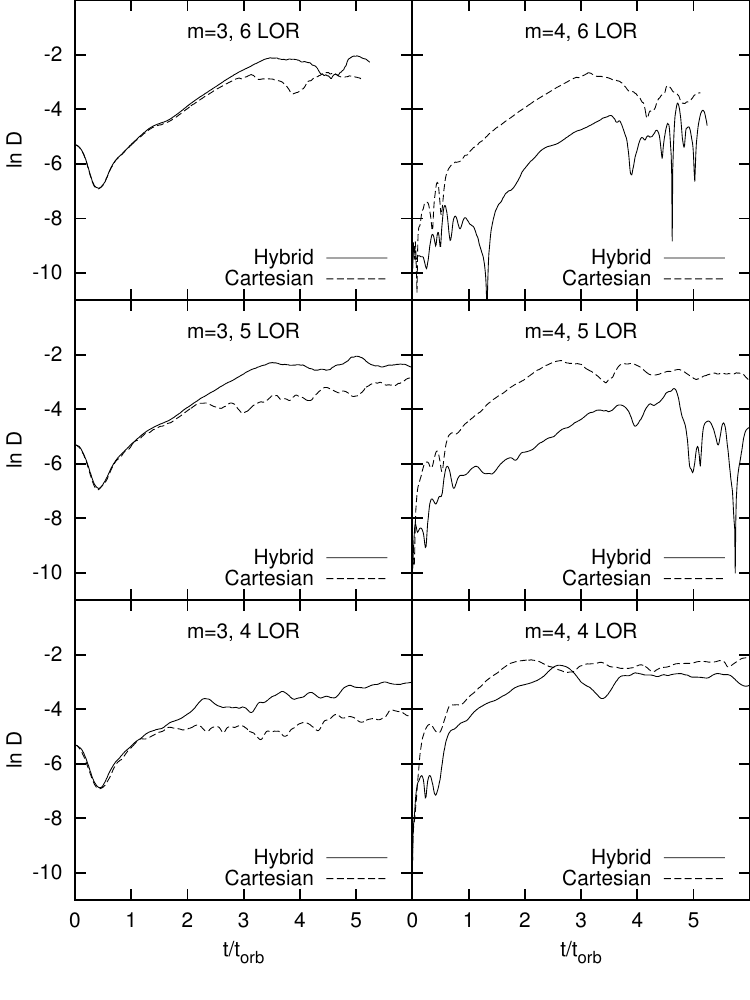}
\caption{The same information as shown in Figure \ref{model_7_compare_LOR}, but grouped differently.  Each frame compares hybrid (solid curve) to Cartesian (dashed curve) momentum advection schemes, for either $m=3$ (left column) or $m=4$ (right column), and at 4, 5, and 6 levels of refinement (bottom, middle, and top rows, respectively).}
\label{model_7_compare3}
\end{figure}

Figures \ref{model_3_compare_LOR} - \ref{model_7_compare3} present data from two initial models, each evolved in 6 separate simulations (2 schemes $\times$ 3 LORs). Figures \ref{model_3_compare_LOR} - \ref{model_3_compare3} present data from Model 3 ($R_-/R_+ = 0.3$); Figures \ref{model_7_compare_LOR} - \ref{model_7_compare3} present data from Model 7 ($R_-/R_+ = 0.7$).  In Figures \ref{model_3_compare_LOR} and \ref{model_7_compare_LOR}, each of the six panels shows the $D_m - t$ plot for the fastest growing unstable mode (either $m=1$ or $m=3$, solid curves) and for $m=4$ (dashed curves) from that initial model evolved in a separate simulation.  The plots shown in the left column were produced from simulations evolved with the hybrid scheme; plots in the right column were produced from simulations evolved advecting Cartesian momentum. Figures \ref{model_3_compare_mode} and \ref{model_7_compare_mode} show the same data, but each frame combines all three LOR and separates the plots of the two modes ($m=1$ and $m=4$ for Figure \ref{model_3_compare_mode}, $m=3$ and $m=4$ for Figure \ref{model_7_compare_mode}). Figures \ref{model_3_compare3} and \ref{model_7_compare3} combine the hybrid scheme and Cartesian plots while separating out the different LOR.  

In Figure \ref{model_3_compare_LOR} we see that all six evolutions show virtually identical behavior of the growth of the $m=1$ mode with time. Hence the measured growth rate is quite independent of the chosen advection scheme or selected LOR. However, from Figure \ref{model_3_compare_LOR} we can also see that the time-dependent behavior of the amplitude of the $m=4$ distortion seems to be strongly resolution dependent.  As stated earlier, this likely reflects the 4-fold symmetry of the Cartesian grid structure.  
In Figure \ref{model_3_compare_mode} the two upper panels ($m=1$) show more clearly that results from both the hybrid scheme and Cartesian scheme are nearly identical, with the lower resolution simulations displaying only slightly noisier $D_m - t$ plots.  The lower two panels show how the amplitude of $m=4$ fluctuations decreases dramatically with increasing resolution.
We can see from the left column of Figure \ref{model_3_compare3}, showing $m=1$, the difference between the Cartesian and hybrid scheme plots differs less with increasing resolution, suggesting that they are both converging to the same answer. In the right column, showing $m=4$, we see that at each resolution the hybrid scheme has slightly lower amplitude.

In contrast to the previous set of figures, Figures \ref{model_7_compare_LOR}-\ref{model_7_compare3} (the slimmest torus evolutions) show a dramatic difference between the hybrid scheme and Cartesian momentum advection scheme. In Figure \ref{model_7_compare_LOR}, the differences are quite apparent not only between the left and right columns (hybrid scheme and Cartesian, respectively) but also between each row (different LOR). This is a model in which the most rapidly growing mode should be $m=3$, but development of unphysical $m=4$ distortions can dominate. Note that for the hybrid scheme, $m=4$ dominates only at 4 LOR.  For the Cartesian, $m=4$ dominates at 4 and 5 LOR, and is only matched by $m=3$ at 6 LOR. The increasing amplitude of the $m=3$ mode compared to the $m=4$ mode with resolution is indicative that a dominant $m=3$ is the true character of the most unstable eigenmode for this model.  This is also consistent with the relative amplitudes measured by \cite{kojima1986}, as shown in Figure \ref{y_2comparison}.  
Focusing on the top row of Figure \ref{model_7_compare_mode}, we see that the $m=3$ mode amplitude tracks the amplitude of the next higher level of resolution until a certain point where it turns off (for hybrid 4 LOR, Cartesian 4 and 5 LOR).  Looking below to the $m=4$ plots, we can see that these turn-off points correspond to a time where the $m=4$ amplitude has risen to $\ln D_4 \approx -3$.  Apparently, once a mode reaches this amplitude, the modes are no longer sufficiently decoupled, and the $m=3$ cannot continue to grow exponentially at the rate predicted by simulations that do not allow different modes to couple.  We also observe that the amplitude of the $m=4$ mode decreases with increasing resolution. In Figure \ref{model_7_compare3} we again see dramatic differences between the hybrid scheme and Cartesian momentum scheme, especially in $m=4$. 

For the $R_-/R_+ = 0.7$ model, simulations that relied on Cartesian momentum
advection failed to achieve qualitative convergence even at 6 LOR. So, in order
to show that both schemes (the Cartesian momentum advection and the hybrid
scheme) ultimately converge to the same answer, the model was re-run advecting
Cartesian momentum at 7 LOR.  This is impossible using the OpenMP version
of the code, so the simulation was run using Octopus, which employs computational fluid algorithms identical to those used in Scorpio, but built within the HPX parallel programming \label{octopus}
framework\footnote{http://stellar.cct.lsu.edu/}. The Octopus code exposes a
greater degree of parallelism than the OpenMP code, making the simulation
practical.  Octopus parallelizes the invocation of various ``kernels'' to every
subgrid in the octree-based grid structure.  In contrast, the OpenMP code only
parallelizes local loops inside these kernels while the invocation of the
kernels is done serially. For example, in Octopus, the computation of fluxes
for all subgrids is done in parallel; in the OpenMP code, the computation of
the fluxes contained within each subgrid are computed in parallel, but only one
subgrid's fluxes are computed at a time.

\begin{figure}
\centering
\includegraphics[scale=1.7]{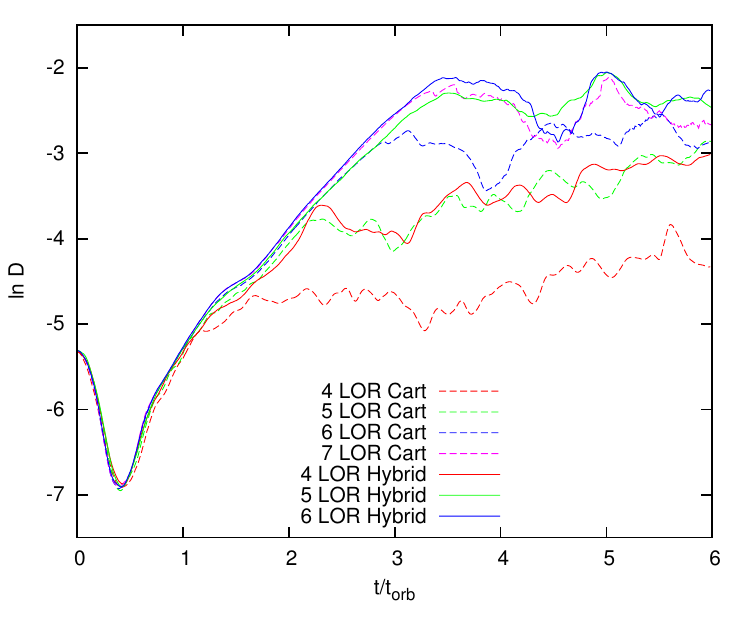}
\caption{
$D_m - t$ plots showing the development of the $m=3$ mode from seven different simulations of Model 7 ($R_-/R_+ = 0.7$). The hybrid scheme simulations (solid curves) show qualitative convergence at 5 LOR, while the Cartesian momentum advection scheme (dashed curves) does not converge until 7 LOR, requiring a factor of $\sim 30$ more computational zones.}
\label{7LOR_compare_3}
\end{figure}

\begin{figure}          
\centering
\includegraphics[scale=2]{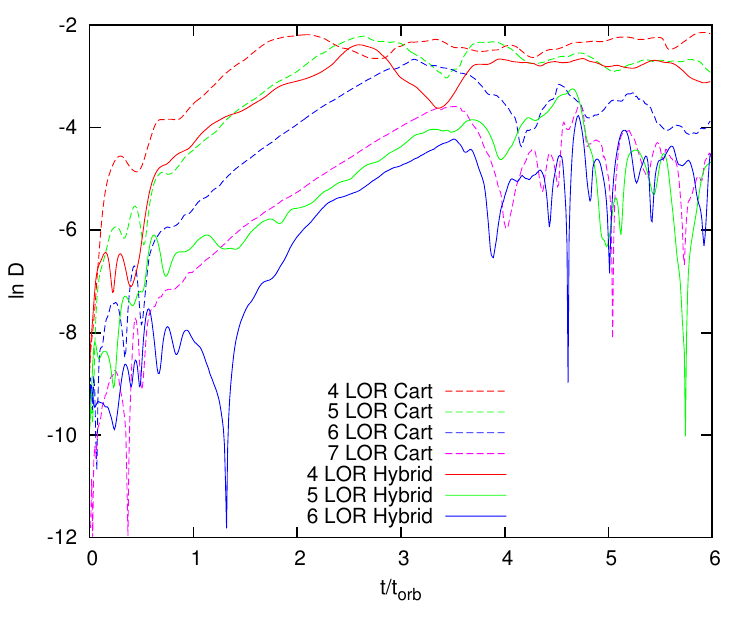}
\caption{
$D_m - t$ plots showing the development of the $m=4$ distortion in Model 7 ($R_-/R_+ = 0.7$). At each specified LOR, the hybrid scheme (solid curves) shows much lower levels of development of this unphysical distortion than the Cartesian advection scheme (dashed curves). 
}
\label{7LOR_compare_4}
\end{figure}

The $D_m - t$ plots presented in Figure \ref{7LOR_compare_3} show the development of the $m=3$ mode from seven separate simulations of Model 7 ($R_-/R_+ = 0.7$). Combining the top two panels of Figure \ref{model_7_compare_mode}, three different resolutions, 4, 5, and 6 LOR, are shown for both the hybrid scheme and the Cartesian momentum advection; additionally, 7 LOR is shown for the Cartesian simulation (red dashed curve).  While the 4, 5, and 6 LOR curves for the Cartesian simulations don't agree with the results of the converged hybrid scheme curves (at 5 and 6 LOR), the 7 LOR Cartesian does lie almost directly on top of those curves.  This demonstrates that both schemes do ultimately converge to the same answer, but that, in this case, the hybrid scheme achieves qualitative convergence a full 2 levels of refinement sooner, and with a factor of $\sim 30$ fewer fine zones.  Figure \ref{7LOR_compare_4} shows the behavior of the $m=4$ distortion for the same set of simulations.  We see that the amplitude of the $m=4$ fluctuation is strongly resolution dependent and, furthermore, that the hybrid scheme demonstrates much lower levels of the unphysical $m=4$ distortion than the Cartesian advection scheme. 

\begin{figure}
\centering
\includegraphics[scale=0.3]{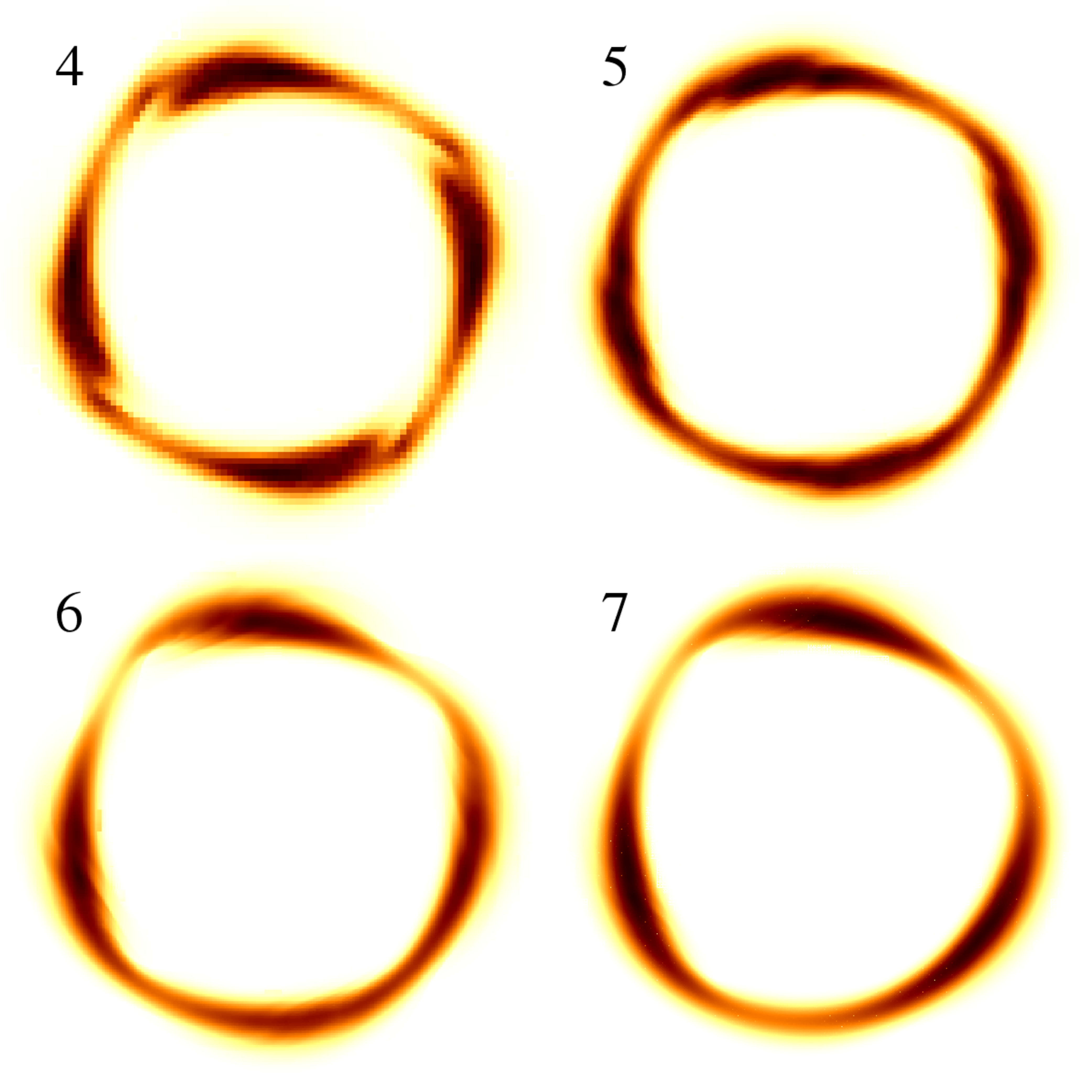}
\caption{Mass density plots from the Cartesian momentum advection simulations of the Model 7 ($R_-/R_+ = 0.7$) torus, showing the progression from $m=4$ dominated evolutions at lower resolutions to $m=3$ dominated evolutions at the highest resolution. Each slice shows the mass density in the $z=0$ plane, and is labeled with the number of LOR used in the simulation. Data for 4, 5, 6, and 7 LOR are taken from $t=1.8t_\mathrm{orb}$, $2.1t_\mathrm{orb}$, $3.0t_\mathrm{orb}$, and $2.5t_\mathrm{orb}$ respectively. Note that the simulations are the same as shown in Figure 2, although from different times in the evolutions.
}
\label{state_data}
\end{figure}

    Figure \ref{state_data} illustrates what the nonlinear amplitude structure of these modes looks like physically.  These equatorial ($z=0$) plane mass density plots drawn from the Cartesian momentum advection simulations confirm what the Figure \ref{7LOR_compare_3} and \ref{7LOR_compare_4} $D_m - t$ plots tell us about the relative amplitudes of the $m=3$ mode and the $m=4$ distortion. Specifically,  at 4 and 5 LOR, the $m=4$ distortion is clearly dominant. At 6 LOR, the $m=3$ mode and $m=4$ distortion are approximately the same amplitude, and the mass density distribution shows this.  At 7 LOR, the mass density distribution shows a dominant $m=3$ mode.

\begin{deluxetable}{ccclllllll}
\tablecolumns{8}
\tablewidth{0pt}
\tablecaption{Torus Mode Characteristics (5 LOR)}
\label{table:results}
\tablehead{\colhead{Model} & \colhead{$m_\mathrm{kick}$} & Scheme\tablenotemark{a} &  \colhead{$\bar{S}$}  & \colhead{$y_2$} & \colhead{$\delta_{y_2}$} & \colhead{$\frac{d \phi}{dt}$} & \colhead{$y_1$} & \colhead{$\delta_{y_1}$} \\
\colhead{(1)} & \colhead{(2)} & \colhead{(3)} & \colhead{(4)} & \colhead{(5)} & \colhead{(6)} & \colhead{(7)} & \colhead{(8)} & \colhead{(9)}
}
\startdata
1 & 1 & C &  0.760 & 0.121  &  0.007  &  2.294  &  0.365  &  0.0006  \\
1 & 1 & H &  0.752 & 0.120  &  0.009  &  2.103  &  0.335  &  0.0005  \\
2 & 1 & C &  0.792 & 0.126  &  0.005  &  1.885  &  0.300  &  0.0002  \\
2 & 1 & H &  0.780 & 0.124  &  0.005  &  1.834  &  0.292  &  0.0004  \\
3 & 1 & C &  1.031 & 0.164  &  0.006  &  1.578  &  0.251  &  0.0004  \\
3 & 1 & H &  0.994 & 0.158  &  0.005  &  1.579  &  0.251  &  0.0003  \\
4 & 1 & C &  1.047 & 0.167  &  0.008  &  1.064  &  0.169  &  0.0005  \\
4 & 1 & H &  1.064 & 0.169  &  0.011  &  1.064  &  0.169  &  0.0002  \\
5 & 2 & C &  1.253 & 0.199  &  0.020  &  1.251  &  0.199  &  0.0017  \\
5 & 2 & H &  1.093 & 0.174  &  0.016  &  1.256  &  0.200  &  0.0027  \\
6 & 2 & C &  1.313 & 0.209  &  0.027  &  0.594  &  0.095  &  0.0011  \\
6 & 2 & H &  1.320 & 0.210  &  0.011  &  0.792  &  0.126  &  0.0007  \\
7 & 3 & C & 1.220 & 0.194  &  0.037  &  0.468  &  0.074  &  0.0017  \\
7 & 3 & H & 1.208 & 0.192  &  0.014  &  0.553  &  0.088  &  0.0005 \\
\enddata
\tablenotetext{a}{H denotes hybrid advection scheme, C denotes Cartesian momentum advection.}
\end{deluxetable}

\section{Conclusion}

Our ultimate goal is to model in as realistic a manner as possible the dynamical evolution of mass-transferring binary systems. This can only be accomplished if the hydrodynamic code that is used to perform each simulation conserves angular momentum extremely well.  We also need to have the flexibility of AMR to adequately resolve spatial features across many orders of magnitude in length scales simultaneously. The hybrid scheme described here allows us to conserve angular momentum to high accuracy on a refined Cartesian mesh, facilitating the use of AMR.

Our hybrid scheme is an implementation of the theoretical formulation developed by \cite{Call2010}, which shows that we have the freedom to choose different coordinate bases for the transport velocity relative to the grid and the advected momentum quantities.  In the past, these chosen basis sets typically have been the same -- resulting in the advection of Cartesian momentum components on a Cartesian mesh, or cylindrical momentum components on a cylindrical mesh. In the hybrid scheme implemented here, we have chosen to advect cylindrical momentum components across a rotating Cartesian mesh.  This allows us to conserve angular momentum to machine precision while capitalizing on the advantages of a Cartesian mesh, such as mesh refinement. 

In order to test this method, we followed the development of nonaxisymmetric instabilities in massless PP tori having $n=3$ and $q=2$ (uniform specific angular momentum). This is a well-defined, fully three-dimensional problem with a reproducible solution.  We evolved seven different initial tori with aspect ratios ranging from $R_-/R_+=0.1$ to $0.7$.  We chose to evolve two particular models, $R_-/R_+ = 0.3$ and $0.7$, using several different grid resolutions.  We compared our results to the linear stability analysis of \cite{kojima1986} and to the nonlinear hydrodynamics results of \cite{WTH1994}.  Our code achieved good agreement with results from these previous studies.

We also introduced a prescriptive method for measuring the real and imaginary parts of the eigenfrequency of unstable modes, attaching an uncertainty to those measurements.  This was done in an effort to increase transparency, reduce the influence of human judgment, and facilitate the reproducibility of these simulations. Through this work we have illustrated the utility of the PP tori as a new test problem, to be added to the standard suite of hydrodynamic test problems, that provides a means for measuring the ability of a particular code to correctly transport and conserve angular momentum. 

A comparison of the resolution dependence of the hybrid scheme compared to the Cartesian momentum advection scheme shows that the hybrid scheme achieves qualitative convergence at grid resolutions that are equal to or lower than the Cartesian scheme.  Specifically, we observe that in the $R_-/R_+ = 0.7$ torus, the hybrid scheme achieves qualitative convergence at only 5 LOR, whereas the Cartesian advection scheme required 7 LOR to achieve the same convergence --  requiring a factor of $\sim 30$ more computation zones. The hybrid scheme also reduces the level of unphysical $m=4$ distortions that characteristically appear in simulations involving angular motion across a Cartesian grid.

Here we have demonstrated the utility of the hybrid scheme, which is only one very specific implementation of the  formalism presented by \citet{Call2010}, which can be applied in a fully relativistic generalized coordinate system. 

We acknowledge valuable interactions that we have had with J. Frank, G. Clayton, P. M. Motl, and H. Kaiser over the course of this project. This work has been supported, in part, by grants ACI-1246443 and AST-1240655 from the U.S. National Science Foundation, in part, by NASA ATP grant NNX10AC72G, and, in part, by U.S. Department of Energy grant DE-SC0008714. This research also has been made possible by grants of high-performance computing time at HPC@LSU through the allocation hpc\_dwd\_amr, and across LONI (Louisiana Optical Network Initiative), especially award loni\_lsuastro11.

\appendix

\section{Momentum Equation Formulation}
\label{appendix_eqs}

This appendix is meant to more completely explain the origin of the momentum equations used in this work, and their place in the larger context of astrophysical hydrodynamics codes.  The theoretical justification for the equations we use comes from \cite*{Call2010}, which describes a generalized version of the hybrid advection scheme used in this work, expanded to general relativity and to any curvilinear coordinate system.  The equations used in this work differ from those typically seen in the community because they exploit two key advantages of the formalism described by \cite*{Call2010}:

\begin{itemize}
\item We are allowed to advect inertial-frame quantities on a rotating grid.
\item We are allowed to choose different coordinate bases for the advected momentum quantities and for the grid on which we choose to advect these quantities.
\end{itemize}

Specifically, in the hybrid scheme as implemented here, we choose to advect cylindrical momentum quantities (that is, radial momentum, angular momentum, and $z$-direction momentum) measured in the inertial frame on a Cartesian mesh rotating with a fixed angular velocity.  Below we will show how this simplifies and in certain cases eliminates the source terms from individual components of the momentum equation.

\subsection{Statements of Conservation}
We typically encounter hyperbolic PDEs of the following form:
\begin{equation}
\frac{d}{dt}\Psi + \Psi\nabla\cdot\vc{u} = S,
\end{equation}
where $\vc{u}$ is the velocity of the fluid as viewed from an inertial frame of reference, and the total time derivative is the Lagrangian derivative, following an individual fluid element as it moves through space. When the source term $S$ is zero, then $\Psi$ represents the volume density of a conserved quantity. Mass density, $\rho$, for example, is a conserved quantity, and the continuity equation does in fact have the form:
\begin{equation}
\frac{d}{dt}\rho + \rho\nabla\cdot\vc{u} = 0.
\end{equation}
We should then expect that in the case of an axisymmetric distribution of fluid moving in an axisymmetric potential, when the azimuthal component of the angular momentum is conserved, that we will encounter an equation of the form,
\begin{equation}
\frac{d}{dt}(\rho R u_\varphi) + (\rho R u_\varphi)\nabla\cdot\vc{u} = 0,
\end{equation}
where $R$ is the cylindrical radius, and $u_\varphi = R\dot{\varphi}$ is the azimuthal component of the inertial velocity field $\vc{u}$.

\subsection{Rotating Frame of Reference}
In order to reduce the motion of the fluid through the computational grid (thereby reducing the effects of numerical diffusion and artificial viscosity), we often wish to view the fluid from a rotating reference frame. Mathematically, we will accomplish this by changing the velocity in the divergence term to account for the frame velocity field, that is, we will replace $\vc{u}$ with,
\begin{equation}
\vc{u'} = \vc{u} - \vc{u_\mathrm{frame}}.
\end{equation}
 If the velocity field, $\vc{u_\mathrm{frame}}$, is divergence-free, then the transformation is trivial.  For a frame rotating with angular velocity $\Omega_0$, 
\begin{equation}
\vc{u_\mathrm{frame}} = R\Omega_0\vc{\hat{e}}_\varphi,
\end{equation}
and, utilizing cylindrical coordinates,
\begin{equation}
\nabla\cdot\vc{u_\mathrm{frame}} = \frac{\partial}{\partial R}(0) + \frac{1}{R}\frac{\partial}{\partial\varphi}(R\Omega_0) + \frac{\partial}{\partial z}(0) = 0.
\end{equation}
Hence,
\begin{equation}
\frac{d}{dt}\Psi + \Psi\nabla\cdot\vc{u'} = \frac{d}{dt}\Psi + \Psi\nabla\cdot[\vc{u}-\vc{u_\mathrm{frame}}] =  \frac{d}{dt}\Psi + \Psi\nabla\cdot\vc{u}
\end{equation}
so the new hyperbolic PDE becomes,
\begin{equation}
\frac{d}{dt}\Psi + \Psi\nabla\cdot\vc{u'} = S,
\end{equation}
and we are confident that this new PDE represents the physics of the system as well as the original PDE.

\subsection{Eulerian Representation}
In order to follow the time-rate of change of a quantity with respect to a point in space fixed with respect to the chosen frame of reference, we must use the following transformation from the Lagrangian to the Eulerian representation:
\begin{equation}
\frac{d}{dt}\Psi \rightarrow \frac{\partial}{\partial t}\Psi + \vc{u'}\cdot\nabla\Psi.
\end{equation}
We can then rewrite the hyperbolic PDE as,
\begin{equation}
\frac{\partial}{\partial t}\Psi +\vc{u'}\cdot\nabla\Psi +\Psi\nabla\cdot\vc{u'} = S
\end{equation}
or, more succinctly,
\begin{equation}
\label{rotating_frame_equation}
\frac{\partial}{\partial t}\Psi + \nabla\cdot(\Psi\vc{u'}) = S.
\end{equation}
We can recover the inertial-frame version of the equation simply by setting $\Omega_0 = 0$, which is equivalent to setting $\vc{u'}=\vc{u}$,
\begin{equation}
\label{inertial_frame_equation}
\frac{\partial}{\partial t}\Psi + \nabla\cdot(\Psi\vc{u}) = S.
\end{equation}

While the underlying physics is identical, a distinction must be made regarding how the two equations are interpreted. Equation (\ref{inertial_frame_equation}) represents the time-rate of change of $\Psi$ at a fixed point in inertial space, while equation (\ref{rotating_frame_equation}) provides the time-rate of change of $\Psi$ at a fixed point in the rotating coordinate frame.  Notice that this is totally independent of what quantity, $\Psi$, we choose to advect.

\subsection{Angular Momentum Conservation}
When the three vector components of the Euler equation of motion are projected onto a non-rotating cylindrical coordinate grid, the azimuthal component may be written as,
\begin{equation}
\frac{d}{dt}(\rho R u_\varphi) + (\rho R u_\varphi)\nabla\cdot\vc{u} = -\frac{\partial }{\partial \varphi}p - \rho\frac{\partial}{\partial\varphi}\Phi.
\end{equation}
For this equation, the source term is,
\begin{equation}
S = -\frac{\partial }{\partial \varphi}p - \rho\frac{\partial}{\partial\varphi}\Phi,
\end{equation}
and $\Psi = (\rho R u_\varphi)$ is the inertial-frame angular momentum density with respect to the $z$-coordinate axis.  This corresponds to ``Case B ($\eta = 3$)'' in \cite*{Call2010}.  Angular momentum will be conserved locally if the source term, $S = 0$.  This will happen if the azimuthal derivative of the gravitational potential and the azimuthal derivative in the pressure are both zero, or if these two terms balance one another (i.e., $\partial p / \partial\varphi = - \rho \partial\Phi/\partial\varphi$).
Based on the discussion above, it is perfectly valid to view the flow from a rotating frame of reference, in which case the equation is simply,
\begin{equation}
\frac{d}{dt}(\rho R u_\varphi) + (\rho R u_\varphi)\nabla\cdot\vc{u'} = -\frac{\partial }{\partial \varphi}p - \rho\frac{\partial}{\partial\varphi}\Phi.
\end{equation}
We can also rewrite these two equations in their Eulerian form,
\begin{equation}
\label{eulerian_nonrotating}
\frac{\partial}{\partial t}(\rho R u_\varphi) + \nabla\cdot[(\rho R u_\varphi)\vc{u}] = S,
\end{equation}
and, when we want to follow the fluid on the rotating coordinate grid,
\begin{equation}
\label{eulerian_rotating}
\frac{\partial}{\partial t}(\rho R u_\varphi) + \nabla\cdot[(\rho R u_\varphi)\vc{u'}] = S.
\end{equation}
When comparing equations (\ref{eulerian_nonrotating}) and (\ref{eulerian_rotating}), notice that the conserved quantity is the same -- the $z$-component of the angular momentum measured in the \textit{inertial} frame. The only difference in the two equations is the ``transport'' velocity ($\vc{u}$ for the nonrotating frame, $\vc{u}'$ for the rotating reference frame).

Equation (\ref{eulerian_rotating}) is different from the more familiar formulation, where the angular momentum density as well as the transport velocity is measured with respect to the rotating frame, i.e., where the angular momentum density is expressed in terms of the azimuthal component of the transport velocity, $u'_\varphi$.  But, as a consequence, the source term in the more familiar formulation is more complicated.  We can derive the more familiar formulation from equation (\ref{eulerian_rotating}) by recognizing that,

\begin{equation}
u_\varphi = u'_\varphi + R\Omega_0.
\end{equation}
So we can write,
\begin{equation}
\frac{\partial}{\partial t}[\rho R(u'_\varphi + R\Omega_0)] + \nabla\cdot\{[\rho R(u'_\varphi + R\Omega_0)]\vc{u'}\} = S_{\varphi i},
\end{equation}
where, as shorthand, we have used,
\begin{equation}
S_{\varphi i} \equiv  -\frac{\partial }{\partial \varphi}p - \rho\frac{\partial}{\partial\varphi}\Phi.
\end{equation}
This implies,
\begin{eqnarray}
\frac{\partial }{\partial t}(\rho R u'_\varphi) + \nabla\cdot[(\rho R u'_\varphi )\vc{u'}] &=& S_{\varphi i} -\frac{\partial}{\partial t}[\rho R(R\Omega_0)] - \nabla\cdot\{[\rho R (R\Omega_0)]\vc{u'}\} \\
&=& S_{\varphi i} - R^2\Omega_0\left\{\frac{\partial}{\partial t}\rho + \nabla\cdot(\rho\vc{u'})\right\} - \rho\vc{u'}\cdot\nabla(R^2\Omega_0)\\
\label{eul_rot_trad}
&=& S_{\varphi i} - 2\rho R u'_R\Omega_0,
\end{eqnarray}
where the last step is accomplished by making use of the continuity relation, $\partial\rho/\partial t + \dv{(\rho\vc{u}')} = 0$. Notice that all velocities now refer to $\vc{u}'$, the velocity as measured in the rotating frame, which is the more familiar formulation. The appearance of a Coriolis term is the result of choosing to measure angular momentum in the rotating frame rather than in the inertial frame. This corresponds to ``Case B ($\eta = 3'$)'' in \cite*{Call2010}.  In our hybrid scheme we have chosen to use equation (\ref{eulerian_rotating}) instead of (\ref{eul_rot_trad}) primarily because equation (\ref{eulerian_rotating}) presents a simpler source term.

\newcommand{\progname}{Scorpio}
\newcommand{\fig}[1]{Figure \ref{#1}}

\newcommand{\algobreak}{
\algstore{myalg}
\end{algorithmic}
\end{algorithm}
\begin{algorithm}
\begin{algorithmic}    
\algrestore{myalg}
}

\section{Code Details}
\label{appendix_code}
The \progname{} mesh structure is an adaptively refined oct-tree of sub-grids. Each sub-grid is composed of an $8\times 8\times 8$ three-dimensional mesh. Each node contains its own sub-grid and up to 
eight child sub-grids. Child sub-grids have one half the grid spacing of their parents. The structure is similar to that used by \cite{MacNeice2000} in PARAMESH, except that a node may have any number of 
children between zero and eight, instead of having either zero or eight children. \fig{octtree} depicts the structure of a simple example mesh with two levels of refinement. In the case shown, the parent 
grid has two of its possible eight child regions refined. One of these child regions is itself refined entirely into eight children, while the other is not refined at all. As with PARAMESH, \progname{} 
requires ``proper nesting"; that is, that there be no more than one jump in refinement across a sub-grid boundary. Referring again to the case depicted in \fig{octtree}, the refinement of child regions 
in the bottom left sub-grid that border the bottom right sub-grid require, by proper nesting, the bottom right sub-grid to exist, regardless of whether or not other refinement criteria are met.  Each sub-grid contains an ``interior" region and a ``ghost" region (3 cells on each side in this work). The interior cells are updated by evolving the solution variables in time while the ghost cells are copied or interpolated from other sub-grids or, in the case of the physical boundaries of the AMR structure, are computed by prescribing an outflow boundary condition. At boundary interfaces between sub-grids of the same refinement level, the ghost cells of a sub-grid are copied from the interior cells of its neighboring sub-grids. Where there is no neighboring sub-grid of the same refinement level and the boundary is not a physical boundary, the ghost cells are computed from the corresponding parent cells. 

The hydrodynamics equations are evolving using the method of \cite{kurganovtadmor} (the ``K-T method" hereafter). The K-T method evolves cell centered quantities without the need for a(n) (approximate) Riemann solver or dimensional splitting. The K-T method requires reconstruction of the evolution variables at left and right cell faces. Any number of reconstruction schemes may be chosen. Scorpio uses the third order piecewise parabolic reconstruction of \cite{CW1984}, without discontinuity detection. The use of a third order reconstruction allows for continuous reconstructions across cell faces in smooth regions of the flow, eliminating the artificial viscosity applied by the K-T method in these regions. The K-T method can be cast in semi-discrete form - discrete in space, but continuous in time. An appropriate time integrator is then used to advance the semi-discrete equations in time. Scorpio uses the explicit $3^\mathrm{rd}$ order total variation diminishing (TVD) Runge Kutta (RK) integrator of \cite{shuosher}. Although both the spatial and temporal discretizations are third order along individual dimensions, because edge and vertex cells are not used in the reconstruction, the overall accuracy reduces to second order.

In Algorithm 1, the algorithm used by \progname{} is presented in a pseudo-code format. A more complete description of \progname{}, along with results from a suite of test problems, is presented by Marcello et al. (2014, in preparation). As has been pointed out in \S2.2, a couple of features of \progname{} that appear in this pseudo-code have not been activated in our present work, namely, AMR and evolving a total energy equation. 

\begin{figure}
\begin{center}
\includegraphics[scale=0.4]{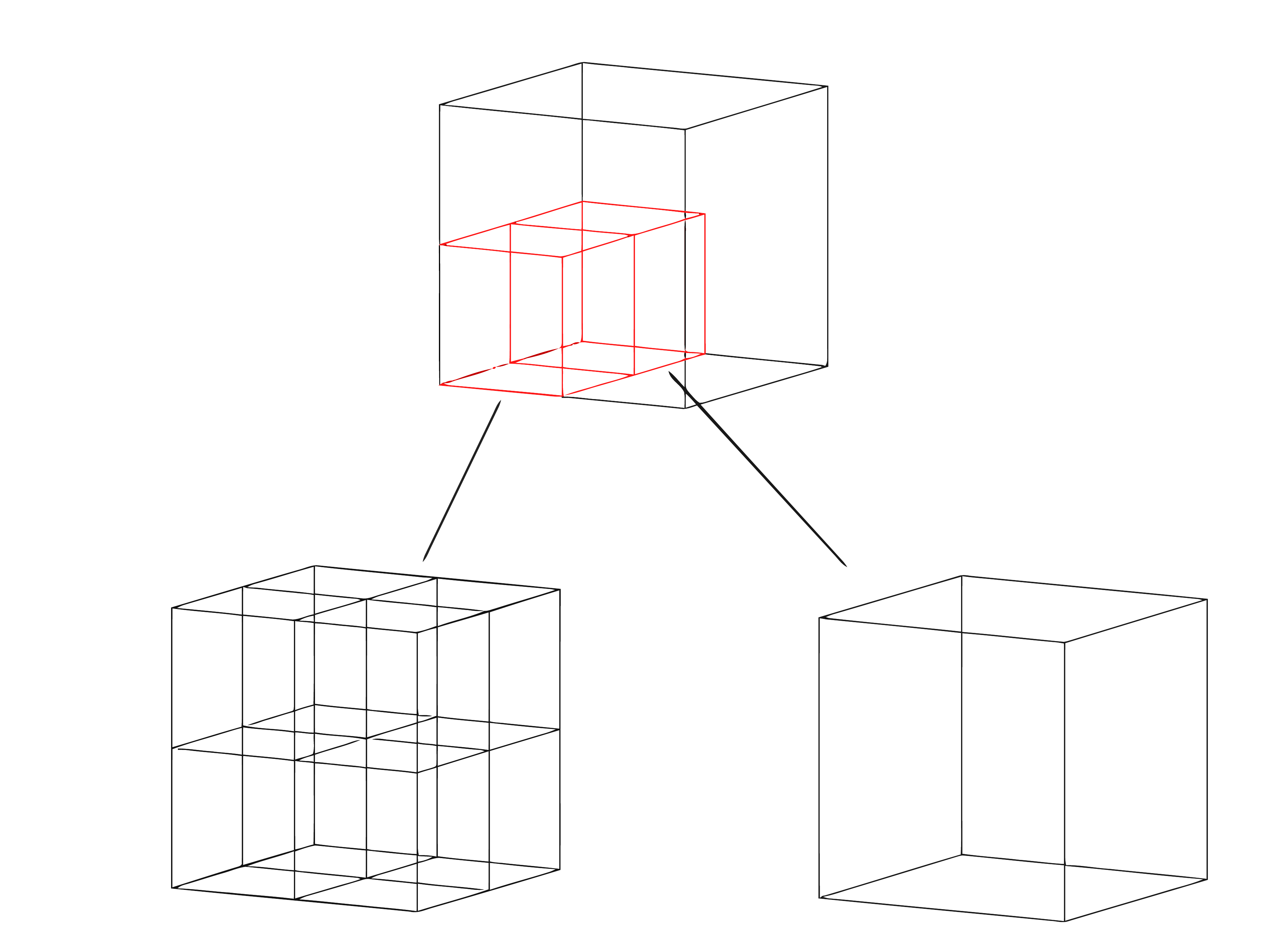} 
\caption{Example oct-tree grid structure. Depicted is a simple two level example mesh. The parent sub-grid (top) is refined in two of its eight child regions. The bottom left is further refined in all eight of its child regions, while the sub-grid on the bottom right is not refined at all. Note that, for added clarity, sub-grids on the bottom are shown at twice their actual size relative to the parent grid at the top.}
\label{octtree}
\end{center}
\end{figure}

\begin{algorithm}
\caption{\progname{} Algorithm}          
\label{salg} 
\begin{algorithmic}
\State $U \gets U\left(t=0\right)$
\State $t \gets 0$ 
\State $n^\mathrm{step} \gets 0$
\While{ $t <= t_\mathrm{max}$}
	\State {Pass decomposition, AMR, and physical boundary values between grids}
	\If{ Boundary is decomposition }
		\State {Take boundary values from neighboring sub-grid}
	\ElsIf{ Boundary is AMR }
		\State {Interpolate boundary values from parent sub-grid}
	\ElsIf{ Boundary is physical }
		\State {Copy outermost interior values to boundary values }
	\EndIf
	\State {Compute primitives $V := \left(\rho, \frac{s_r}{\rho}, \frac{\ell_z}{\rho R} - R \Omega_0, \frac{s_z}{\rho}, E-\frac{1}{2}\rho u^2\right)$ from $U$}
	\State \ \ \ $V \gets U$
	\State {Reconstruct the values of $V$ on cell faces using PPM}
	\State \ \ \ $V_{i\pm\frac{1}{2},j,k}^{r|l}, V_{i,j\pm\frac{1}{2},k}^{r|l}, V_{i,j,k\pm\frac{1}{2}}^{r|l} \gets$ PPM reconstruction of $V$, where the $r$ or $l$ superscripts 
	\State\hspace{\algorithmicindent} refer to the right or left cell faces, respectively.
	\State {Compute CFL condition using signal speeds at cell faces}
	\State $\ \ \ \mathrm{dt} \gets \frac{\alpha_\mathrm{CFL}}{max_V \left\{\lambda_{i+\frac{1}{2},j,k}^{r|l}, 
		\lambda_{i,j+\frac{1}{2},k}^{r|l}, \lambda_{i,j,k+\frac{1}{2}}^{r|l}  \mathrm{dx}\right\}}$,
			where the $\lambda$'s are the signal speeds computed 
			\State\hspace{\algorithmicindent} from the reconstructed values of $V$ at respective cell faces and is a constant 
			\State\hspace{\algorithmicindent} satisfying $\alpha_\mathrm{CFL} \le 0.4$.
	\State {Store the solution vector, $U$, in $U_0$}
	\State \ \ \ $U_0 \gets U$

%        \algobreak
	\For{ $\beta \gets \left\{1,\frac{1}{4}, \frac{2}{3}\right\}$ }
		\State {Compute primitives $V := \left(\rho, \frac{s_r}{\rho}, \frac{\ell_z}{\rho R} - R \Omega_0, \frac{s_z}{\rho}, E-\frac{1}{2}\rho u^2\right)$ from $U$}
		\State \ \ \ $V \gets U$
		\State {Reconstruct the values of $V$ on cell faces using PPM}
		\State \ \ \ $V_{i\pm\frac{1}{2},j,k}^{r|l}, V_{i,j\pm\frac{1}{2},k}^{r|l}, V_{i,j,k\pm\frac{1}{2}}^{r|l} \gets$ PPM reconstruction of $V$.
		\State {Compute face values of $U$ from the reconstructed faces values of $V$}
		\State \ \ \ $U_{i\pm\frac{1}{2} j k}^{r|l} \gets V_{i\pm\frac{1}{2} j k}^{r|l}, 
                              U_{i j\pm\frac{1}{2} k}^{r|l} \gets V_{i j\pm\frac{1}{2} k}^{r|l}, 
                              U_{i j k\pm\frac{1}{2}}^{r|l} \gets V_{i j k\pm\frac{1}{2}}^{r|l} $
		\State {Compute fluxes at cell faces using the PPM reconstructed values and the KT scheme}
		\State \ \ \ 	$F_{i + \frac{1}{2} j k },F_{i  j + \frac{1}{2} k},F_{i  j k + \frac{1}{2}} \gets $ KT fluxes
		\State {Match coarse fluxes to fine fluxes at AMR boundaries. E.g., for x-fluxes,} 
		\State \ \ \ $F_{i_c + \frac{1}{2} j_c k_c }^\mathrm{coarse} \gets 
			\frac{1}{4}\left( F_{i_f+ \frac{1}{2}  j_f - \frac{1}{2} k_f }^\mathrm{fine} + F_{i_f+ \frac{1}{2}  j_f + \frac{1}{2} k_f }^\mathrm{fine} 
			+ F_{i_f+ \frac{1}{2}  j_f  k_f - \frac{1}{2} }^\mathrm{fine} + F_{i_f+ \frac{1}{2}  j_f  k_f + \frac{1}{2}}^\mathrm{fine} \right)$, 
			\\ \ \ \ \ \ \ \ \ \ \ where $\left(i_c + \frac{1}{2} ,j_c,k_c\right)$ and  $\left(i_f+ \frac{1}{2},j_f,k_f\right)$ 
				coincide with an AMR boundary
		\State {Compute the sources terms}
		\State \ \ \ $S_{i j k} \gets$ gravitational and centrifugal source terms
		\algobreak
		\State {Compute the time rate of change for $U$ using the K-T fluxes and source terms} 	
		\State $\ \ \ \dot{U}_{i j k} \gets 
				-\frac{1}{\mathrm{dx}}\left(\left( F_{i + \frac{1}{2} j  k} - F_{i - \frac{1}{2} j k } \right) +
				\left( F_{i  j + \frac{1}{2} k} - F_{i j - \frac{1}{2} k } \right) +
				\left( F_{i  j k + \frac{1}{2}} - F_{i j k - \frac{1}{2} } \right)\right)+ S_{i j k} $
                \State {Update $U$ using RK3 time integrator}
		\State $ \ \ \ U \gets \beta \left(U+\dot{U}\mathrm{dt}\right) + \left(1-\beta\right) U_0$
		\State {Update parent states from children}
		\If {$U_{i_c j_c k_c}$ has children}
			\State $U_{i_c j_c k_c} \gets \frac{1}{8} \sum_{\mathrm{All} \left(i_f j_f k_f\right) \mathrm{in} \left(i_c j_c k_c\right)} U_{i_f j_f k_f}$
		\EndIf
		\State Floor density values 
		\State $\ \ \ \rho \gets \max{\left\{\rho,\rho_\mathrm{floor}\right\}}$
		\State {Pass decomposition, AMR, and physical boundary values between grids}
%                \algobreak
		\If{ Boundary is decomposition }
			\State {Take boundary values from neighboring sub-grid}
		\ElsIf{ Boundary is AMR }
			\State {Interpolate bounty values from parent sub-grid}
		\ElsIf{ Boundary is physical }
			\State {Copy outermost interior values to boundary values}
		\EndIf
	\EndFor

	\State $t \gets t + dt$
	\State $n_\mathrm{step} \gets n_\mathrm{step} + 1$
	\If{ $n_\mathrm{step}  \mod N_\mathrm{RF_\mathrm{freq}} := 0$, where $N_\mathrm{RF_\mathrm{freq}}$ determines how often refinement criteria are
			\State\hspace{\algorithmicindent}  checked. }
		\State {Check for refinement/derefinement}
		\If {Any $\rho_{i j k} \ge \rho_\mathrm{refine}$ in a given sub-grid octant AND $l < l_\mathrm{max}$} ($l$ is 
			\State\hspace{\algorithmicindent} the refinement level and $l_\mathrm{max}$ is the maximum level allowed)
			\State {Mark sub-grid octant for refinement}
		\EndIf
		\If {All $\rho_{i j k} < \rho_\mathrm{refine}$ in a given sub-grid octant} 
			\State {Mark sub-grid octant for derefinement}
		\EndIf
		\If {A given sub-grid octant must be refined to satisfy proper nesting requirements} 
			\State {Mark sub-grid octant for refinement}
		\EndIf
		\State {Create/Destroy sub-grids as needed and initialize new grids using}
		\State \ \ \ $U^\mathrm{coarse} := U^\mathrm{fine}$
	\EndIf
\EndWhile
\end{algorithmic}
\end{algorithm}

\bibliography{paper}

\end{document}